\documentclass[sn-basic,iicol]{sn-jnl}% Basic Springer Nature Reference Style/Chemistry Reference Style
\usepackage{bm}
\usepackage{makecell}

\usepackage[utf8]{inputenc}
\usepackage[capitalize]{cleveref}
\crefname{equation}{}{}

\usepackage{enumitem}

\usepackage{pgfplots}
\pgfplotsset{
  compat=newest,
  label style={font=\small},
  tick label style={font=\small},
  legend style={font=\small, inner xsep=3pt, inner ysep=1pt},
  legend cell align={left},
  scaled y ticks=false,
  /tikz/mark size=2pt
  }
\usepgfplotslibrary{polar}
\usepackage{pgfplotstable}
\usetikzlibrary{arrows}
\usetikzlibrary{fadings}
\usetikzlibrary{shapes}
\usetikzlibrary{arrows.meta}
\usetikzlibrary{calc}

\usepackage{our-macros}
\jyear{2022}%

\theoremstyle{thmstyleone}%
\theoremstyle{thmstyletwo}%
\newtheorem{remark}{Remark}%

\theoremstyle{thmstylethree}%

\begin{document}

\title[Two-Scale Optimization with respect to Buckling]{Two-Scale Optimization of Graded Lattice Structures respecting Buckling on Micro- and Macroscale}

%%=============================================================%%
%% Prefix -> \pfx{Dr}
%% GivenName  -> \fnm{Joergen W.}
%% Particle -> \spfx{van der} -> surname prefix
%% FamilyName -> \sur{Ploeg}
%% Suffix -> \sfx{IV}
%% NatureName -> \tanm{Poet Laureate} -> Title after name
%% Degrees  -> \dgr{MSc, PhD}

\author*[1]{\fnm{Daniel} \sur{H\"ubner}}\email{daniel.huebner@fau.de}

\author[1]{\fnm{Fabian} \sur{Wein}}\email{fabian.wein@fau.de}

\author[1]{\fnm{Michael} \sur{Stingl}}\email{michael.stingl@fau.de}

\affil[1]{\orgdiv{Applied Mathematics, Continuous Optimization}, \orgname{Friedrich-Alexander Universität Erlangen-N\"urnberg}, \orgaddress{\street{Cauerstr. 11}, \city{Erlangen}, \postcode{91058}, \country{Germany}}}


% 150 - 250 words
\abstract{Interest in components with detailed structures increased with the progress in advanced manufacturing techniques in recent years. Parts with graded lattice elements can provide interesting mechanical, thermal, and acoustic properties compared to parts where only coarse features are included. One of these improvements is better global buckling resistance of the component. However, thin features are prone to local buckling. Normally, analyses with high computational effort are conducted on high-resolution finite element meshes to optimize parts with good global and local stability. Until recently, works focused only on either global or local buckling behavior. We use two-scale optimization based on asymptotic homogenization of elastic properties and local buckling behavior to reduce the effort of full-scale analyses. For this, we present an approach for concurrent local and global buckling optimization of parameterized graded lattice structures.
%We present a novel approach to reduce computational cost in design of graded lattice structures, which are optimized with regard to both local and global buckling.
%Based on asymptotic homogenization of elastic properties and local buckling behavior, two-scale optimization is performed.
%The choice of a parameterization for the lattice allows for decoupling of the two scales and precomputation of upscaled properties, which can be reused in various optimization problems.
It is based on a worst-case model for the homogenized buckling load factor, which acts as a safeguard against pure local buckling. Cross-modes residing on both scales are not detected. We support our theory with numerical examples and validations on dehomogenized designs, which show the capabilities of our method, and discuss the advantages and limitations of the worst-case model.}

% 4 - 6 keywords



\maketitle

\section{Introduction}%
The ongoing progress in additive manufacturing allows structures with fine details to gain increasing focus.
Lattice structures in particular, both homogeneous and graded, are utilized in many applications, e.g., thermal management, energy absorption, noise reduction, biomedical engineering, etc. \citep{rahman2022density}.
Lattice infill is also recognized as potentially increasing global buckling resistance of a component \citep{clausen2016exploiting}.
However, fine features are prone to local buckling \citep{ferrari2019revisiting}.

Two-scale optimization \citep{wu2021topology} can be used to design such structures without the need to resolve all the fine details of the full design in a single setting.
The idea of this approach started with the work of \cite{bendsoe1988generating}, in which the design process is divided into two scales:
the macroscopic scale, which describes the overall component, and the microscopic scale, which shows the fine details.
Bends{\o}e and Kikuchi bridged the gap between these two scales by asymptotic homogenization \citep{allaire1997shape}.
This technique yields approximate material properties of microstructures on the macroscopic scale, which can then be used in macroscopic constitutive equations.
Choosing a parameterized microstructure, Bends{\o}e and Kikuchi were able to effectively decouple the scales:
Prior to any optimization procedure a discrete subset of the parameter space is chosen and homogenization is conducted for each of the microstructures gained from this parameter set.
The obtained properties are then interpolated in the continuous parameter space and the interpolated material model can later be reused to solve various optimization problems.
This technique introduces an interpolation error, but requires less computational effort when compared with on the fly homogenization, i.e., homogenization performed for each finite element in the discretized design domain and for each update step of the design during an iterative optimization procedure.
Moreover the interpolation error can be controlled in a rather straight forward way during preprocessing.

Though there is exhaustive literature on optimal design considering the buckling behavior of structures using beam models (\cite{ferrari2019revisiting} and references therein), only a relatively small number of publications for continuum models exist.
The initial problems evolved around finding optimal cross-sections for columns of fixed length and weight subject to uniaxial compression loads \citep{clausen1851form, keller1960shape, tadjbakhsh1962strongest, huang1968optimal, khot1976optimum}.
\cite{neves1995generalized} were the first to conduct topology optimization with respect to buckling based on the method of Bends{\o}e and Kikuchi described above.
However, they encountered various obstacles in the linear buckling analysis, which is stated as an eigenvalue problem \citep{bendsoe2003topology}.
This includes localized modes in low density regions and clustering of eigenvalues when approaching the final design \citep{seyranian1994multiple}.
Actions to alleviate issues with artificial, low-density modes have been proposed, e.g., different interpolation schemes for the stiffness and geometric stiffness matrices \citep{pedersen2000maximization, bendsoe2003topology} or applying an eigenvalue shift based on the last iteration and identifying artificial modes by their contribution to the total strain energy \citep{gao2015topology}.
There are also methods for avoiding low-density regions like filtering and projection of the pseudo-density \citep{larsen2018optimal}, penalization of intermediate values in the objective \citep{allaire1993numerical,allaire1993explicit}, or element removal strategies \citep{behrou2021revisiting,dalklint2020eigenfrequency,giele2021approaches}.
Clustering of eigenvalues can be prevented by enforcing gaps between the eigenvalues \citep{bendsoe2003topology}.
However, a large number of eigenvalues may still have to be computed to achieve good convergence \citep{bruyneel2008discussion}, which compels the use of efficient eigenproblem solvers \citep{dunning2016level,ferrari2020towards}.
Nevertheless, topology optimization with respect to buckling still currently poses a challenging problem.

More recently, stability requirements have also been employed when tailoring microstructures \citep{neves2002topology,neves2002topologycriteria,thomsen2018buckling,andersen2022buckling}, though models for the buckling of periodic microstructures have been investigated for decades.
Homogenization theory for buckling load factors is well established \citep{neves2002topology,thomsen2018buckling}, but several challenges still arise in this context.
Buckling modes can range from high-frequency modes with a wavelength shorter than the characteristic size of the microstructure to modes that span over multiple periods of the microstructure.
Floquet-Bloch theory can be used to capture the latter in particular \citep{neves2019symbolic}.

For dehomogenized designs scale effects stemming from a finite cell size and effects due to grading of the microstructure might appear \citep{thomsen2018buckling}.
Especially the latter might lead to an occurrence of undesired buckling modes at the boundary of the structure.

The aforementioned works only investigate the buckling behavior of a structure on a single scale, either macroscopic or microscopic.
Even with the ongoing "competition for ultimately stiff and strong architected materials" \citep{andersen2021competition}, these designed materials appear not to have been utilized so far in the optimization of components with stability requirements.

In this article, we present a two-scale topology optimization approach, where we incorporate buckling on both scales individually, but neglect cross-modes spanning over both scales.
Assuming separation of length scales, we use homogenization theory to upscale the elastic behavior and buckling stability of a periodic lattice that resides on the microscopic scale.
On the macroscopic scale we search for an optimal lattice grading with respect to both local and global buckling.
We parameterize the lattice by its porosity, precompute homogenized properties for a selected set of porosities, and obtain a model on the continuous parameter space by piecewise cubic interpolation.
This model is then used in macroscopic optimization problems with the local lattice porosity as design variable.
Thus, our method can be classified as a multi-scale density optimization approach with a parameterized unit cell based on the microstructure porosity (category V/A in \cite{wu2021topology}).

Our contribution is a new approach to integrate the upscaled microscopic buckling information into the optimization.
From the macroscopic local stress, we obtain the buckling response on the microscale via a worst-case model, which acts as a safeguard against pure microscopic buckling.
The worst case is realized by reducing the macroscopic local stress to its magnitude for reasons of computational effort.
This comes at the cost of underestimation of predicted critical loads for the microstructure.

Recent work by \cite{wang2021optimization} is close to our approach.
However, they deduce local stress constraints from the slenderness ratio of the lattice struts, while we directly use upscaled microscopic buckling information.
Our approach has the advantage that essentially arbitrary microstructures can be treated, including those for which a slenderness ratio could not be uniquely defined.

The work of \cite{christensen2022multiscale} is similar to ours, with some key differences:
They assume isotropic buckling behavior of the microstructure, while our approach is applicable to arbitrary microstructures.
They also fit a Willam-Warnke failure criterion to homogenized data, which introduces an approximation error of the local buckling factor.
In contrast, we build a worst-case model from homogenized data, which also comes with an approximation error;
however, applying a full interpolation model in lieu of the worst case, it is possible to render this error negligible (see \cref{rem:wcfree} in \cref{ssec:worstcasemodel}).
Moreover, \cite{christensen2022multiscale} employ a two term interpolation scheme to interpolate the buckling strength as a function of the relative porosity, whereas we use more accurate piecewise cubic interpolation.
On the other hand, our approach is limited to sizing problems with a lower bound on the lattice density everywhere, while a major contribution of \cite{christensen2022multiscale} is the possibility to allow for void regions in the design while still being able to maintain the lower bound on the density of the lattice using an auxiliary design field.

In this article, we restrict the analyses to linear elasticity and linearized buckling, though the general idea of the method can be extended to non-linear regimes.
We also limit ourselves to a two-dimensional setting in this article to keep notations simple.
However, it is straightforward to extend our approach to three dimensions.
We emphasize that our method is applicable to arbitrary, parameterized microstructures.
As an example, we choose a lattice consisting of equilateral triangles that is parameterized by its porosity.
More sophisticated geometries and parameterizations are possible with higher computational effort without changing the general method.

The remainder of the article is structured as follows:
In \cref{sec:linela} we recap the state equations for linear elasticity and linearized buckling analysis.
\cref{sec:homogenization} briefly presents the main formulas for asymptotic homogenization of elasticity modulus and buckling load factors.
We describe our exemplary microstructure, its parameterization, and important aspects for the upscaling process in \cref{sec:parameterization}.
We present our method for designing a lattice unit cell by rounding corners of an equilateral triangle, showing finite element convergence on the microscopic scale, and obtaining homogenized properties of a lattice.
\cref{ssec:worstcasemodel} provides the novel approach to include the upscaled properties on the macroscopic scale via the suggested worst-case model for the homogenized buckling load factor.
\cref{ssec:scaledecoupling} demonstrates how to precompute homogenized properties on a discretized parameter space and apply an interpolation scheme to yield values on the continuous parameter space.
\cref{sec:sizingopt} outlines three different two-scale optimization problems considering buckling on the macroscopic scale, on the microscopic scale and on both scales (without cross-modes).
Associated numerical examples are presented in \cref{sec:numericalexamples} together with a pre-study, which helps to develop a better understanding of the optimized designs.
A numerical validation on dehomogenized designs is given in \cref{ssec:dehomogenization} accompanied by a study on scale separation.
We examine the impact of the worst-case error closely in \cref{ssec:worstcaseimpact} by means of two numerical examples.
Finally, we complete with conclusions in \cref{sec:conclusion}.

\section{Linear elasticity and buckling analysis}\label{sec:linela}%
In this section we briefly recap the state equations for linear elasticity and buckling analysis on an algebraic level as given by \cite{thomsen2018buckling}.
For continuous formulations in weak form, interested readers are referred to \cite{neves2019symbolic}. 
We assume a linear elasticity setting and linear bifurcation buckling condition. This means that the prebuckling displacement, stresses, and strains vary linearly with the applied load and that the load factor, which indicates stability, appears linearly in the bifurcation eigenvalue problem.
Linear buckling analysis consists of two steps:
First, we solve the linear elasticity state equation for displacements of a structure under a given load.
Then, an eigenvalue problem is solved, where the eigenvalues correspond to bifurcation points, and eigenvectors are interpreted as buckling modes of deflection.

Throughout this article, we apply Voigt notation for tensors and assume plane stress conditions.

Let us consider a body $\Omega\subset\RR^2$ and its discretization $\Omega_\vartriangle=\bigcup_{e=1}^\text{M} \Omega_e$ by $\text{M}$ finite elements.
Then the state equation of linear elasticity for $\Omega_\vartriangle$ reads as \citep{zienkiewicz2005finite}:
\begin{align}
  \vec K(\vec\rho)\vec u &= \vec f, \label{eq:linEla}
\end{align}
where $\vec u$ is the sought-for vector of displacements, $\vec\rho\in(0,1]^\text{M}$ is the pseudo-density field and $\vec f$ is the applied load vector (reference load).
The stiffness matrix is given by
\begin{equation}\label{eq:stiffnessmatrix}
\vec K(\vec\rho) = \ass_{e=1}^\text{M} \sum_{k=1}^{\Nip} c_{e,k} \vec B_{e,k}^\top \vec E(\rho_e) \vec B_{e,k}
\end{equation}
with assembly operator denoted by $\ass$ and number of integration points $\Nip$.
$\vec E(\rho_e)$ is the elasticity tensor, which depends on the pseudo-density in element $e$.
$\vec B_{e,k}$ is the strain-displacement matrix of element $e$ evaluated in the $k$-th integration point and contains derivatives of the finite element's shape functions.
The factor
\begin{equation}
c_{e,k} = \det(\vec J_e) w_{e,k}
\end{equation}
contains the Jacobian determinant $\det(\vec J_e)$ of element $e$ and the integration weight $w_{e,k}$ associated with the $k$-th integration point of said element.

The buckling equation is given by the eigenvalue problem,
\begin{align}\label{eq:macrobuckling}
  \left(\vec K(\vec\rho)-\Lambda\, \vec G(\vec\rho,\vec u(\vec\rho))\right)\vec\phi &= 0,
\end{align}
and is solved for pairs of eigenvalues $\Lambda_\ell$ and eigenvectors $\vec\phi_\ell$, $\ell\leq \text{M}$.
Eigenvalues of \cref{eq:macrobuckling} are also called load factors.
The critical load, under which buckling occurs, can be computed by multiplying the reference load $\vec f$ with the eigenvalue with the smallest absolute value, which is also known as the critical load factor or buckling load factor (BLF). 
If the critical load factor is less than or equal to one, the structure buckles; otherwise, the structure is stable with respect to the applied reference load.
The stress stiffness matrix (or geometric stiffness matrix) \citep{nvemec2016new} in \cref{eq:macrobuckling} is given by
\begin{align}\label{eq:stressmatrix}
&\vec G(\vec\rho,\vec u(\vec\rho)) = \notag\\
&\ass_{e=1}^\text{M} \sum_{k=1}^{\Nip} c_{e,k} \vec{\tilde{B}}_{e,i,k} (\vec\sigma_e(\vec u_e(\vec\rho)))_{i,j} \vec{\tilde{B}}_{e,j,k},
\end{align}
where summation over $i$ and $j$ is implied.
Again, $c_{e,k}$ includes the Jacobian determinant of element $e$ and the integration weight at the $k$-th integration point.
The integrated stress in element $e$ is numerically evaluated as
\begin{equation}\label{eq:macrostress}
\vec\sigma_e(\vec u_e(\vec\rho)) = \sum_{k=1}^{\Nip} c_{e,k} \vec E(\rho_e) \vec B_{e,k} \vec u_e(\vec\rho),
\end{equation}
and
\begin{equation}
\vec{\tilde{B}}_{e,i,k} = \tfrac{\partial\vec N_e}{\partial x_i}(x^k)
\end{equation}
is the derivative of the matrix of the shape functions evaluated at integration point $x^k$.
Usually, solutions to \cref{eq:macrobuckling} are considered ordered according to the absolute value of the load factors with $\Lambda_1$ as the smallest value and eigenvectors are $\vec G$-normalized, i.e.,
\begin{equation}
\vec\phi^\top \vec G\, \vec\phi = 1.
\end{equation}
In single-scale topology optimization, \cref{eq:macrostress} can lead to artificial buckling modes in low density regions \citep{neves1995generalized}.
However, in our approach the macroscopic pseudo-density represents the local relative lattice volume.
Thus, low density regions represent thin lattice, and modes in such regions are not artificial, but rather comply with buckling of the lattice.

Following \cite{rodrigues1995necessary}, the sensitivity of an eigenvalue with algebraic multiplicity one with respect to density $\rho_e$ is given  as
\begin{equation}\label{eq:macrobucklingderivative}
 \frac{\partial\Lambda}{\partial\rho_e}
  = \left[\vec\phi^\top\left(\frac{\partial\vec K}{\partial\rho_e} - \Lambda\frac{\partial\vec G}{\partial\rho_e}\right)\vec\phi - \Lambda\vec w^\top \frac{\partial\vec K}{\partial\rho_e}\vec u\right].
\end{equation}
Here, $\vec w$ is the adjoint state obtained from the solution of the equation
\begin{equation}
  \vec K(\vec\rho)\,\vec w = \vec\phi^\top \nabla_u\vec G(\vec\rho, \vec u(\vec\rho))\,\vec\phi.
\end{equation}
For eigenvalues with multiplicity greater one, the derivative of the load factor function does not exist in a strict sense. Nevertheless, using \cref{eq:macrobucklingderivative} still an element of the subdifferential can be computed \citep{rodrigues1995necessary}.
Alternatively, derivations for multiple eigenvalues can be used \citep{seyranian1994multiple}.

Note that $\vec G$ is in general not positive definite, so sometimes a reformulation of the eigenvalue problem \cref{eq:macrobuckling} to

\begin{equation}
  \left(\vec G(\vec\rho,\vec u(\vec\rho))-\tfrac{1}{\Lambda}\vec K(\vec\rho)\right)\vec\phi = 0,
\end{equation}
might be beneficial, as $\vec K$ is always positive definite for $\vec\rho>0$.

\section{Asymptotic homogenization}\label{sec:homogenization}%
To obtain the mechanical properties of a lattice structure on the macroscopic scale, we perform asymptotic homogenization.
We briefly recall the homogenization formulas in the two-dimensional setting, which can be found, e.g., in \cite{thomsen2018buckling}.
Note that we do not perform topology optimization on the microscopic scale, but rather investigate given periodic lattice structures.
Thus, there is only a finite element mesh for the solid material region.

This prevents spurious modes, which typically arise in low density regions during topology optimization \citep{neves1995generalized}.
However, we might get artificial modes owing to the finite element discretization.
These modes are usually highly localized and can be filtered by setting a threshold on the minimal number of nodes that have to exhibit deflection, e.g., $5\%$ of all nodes in the finite element mesh.

As we do not perform topology optimization on the microscopic scale and are only interested in obtaining the critical load factor of a given microstructure, sensitivities are not required here and thus clustering of eigenvalues as described in the work by \cite{neves2002topologycriteria} is not an issue.

\subsection{Linear elasticity}
We choose a representative volume element (RVE) $Y$, which is discretized with $\text{m}$ finite elements $Y_e$, i.e., $Y=\bigcup_{e=1}^\text{m} Y_e$.
We solve the equilibrium equations on the microscopic scale for the $Y$-periodic fields $\vec\chi^j$ using three test fields $\vec f^j$:
\begin{equation}\label{eq:linelahom}
  \vec K \vec\chi^j = \vec f^j, \qquad j=1,2,3
\end{equation}
with stiffness matrix $\vec K$ and $\vec f^j$ given by
\begin{equation}
  \vec K = \ass_{e=1}^\text{m} \sum_{k=1}^{\nip} c_{e,k} \vec B_{e,k}^\top\vec E^0\vec B_{e,k}
\end{equation}
and
\begin{equation}
  \vec f^j = \ass_{e=1}^\text{m} \sum_{k=1}^{\nip} c_{e,k} \vec B_{e,k}^\top\vec E^0 \vec{\tilde\epsilon}^j.
\end{equation}
$\nip$ is the number of integration points, $\vec E^0$ is the elasticity tensor for the base material and $\vec{\tilde\epsilon}^j$ are unit strain fields with $\tilde\epsilon_i^j = \delta_{ij}$. We note that the $Y$-periodicity is realized by assuming same values for the solutions on opposite sides of $Y$.
The homogenized constitutive tensor $\vec E^H$ of the microstructure is then given by
\begin{align}\label{eq:homtensor}
  &\vec E^H_{i,j} = \notag\\
  &\frac{1}{|Y|} \ass_{e=1}^\text{m} \sum_{k=1}^{\nip} c_{e,k} \left( \vec{\tilde\epsilon}^i - \vec B_{e,k}\vec\chi_e^i \right)^\top \vec E^0 \left(\vec{\tilde\epsilon}^j - \vec B_{e,k}\vec\chi_e^j \right).
\end{align}

\subsection{Buckling}
The equilibrium equation for buckling on the microscopic scale is an eigenvalue problem as in the macroscopic setting (\cf \cref{eq:macrobuckling}) and is solved for pairs of eigenvalues $\lambda_\ell$ and associated eigenvectors $\vec\varphi_\ell, \ell\leq m$:
\begin{equation}\label{eq:microbuckling}
  (\vec K-\lambda\, \vec G(\vec\chi,\vec{\bar\sigma}))\vec\varphi = 0
\end{equation}
As in the macroscopic setting, the eigenvalue with the smallest absolute value denotes the buckling load factor.
$\vec G$ is the microscopic initial stress stiffness matrix (\cf \cref{eq:stressmatrix})
\begin{align}\label{eq:stressmatrixmicro}
  &\vec G(\vec\chi,\vec{\bar\sigma}) = \notag\\
  &\ass_{e=1}^\text{m} \sum_{k=1}^{\nip} c_{e,k} \vec{\tilde{B}}_{e,i,k}^\top (\vec\sigma_e(\vec\chi,\vec{\bar\sigma}))_{i,j} \vec{\tilde{B}}_{e,j,k}
\end{align}
with
\begin{equation}\label{eq:stressdistribution}
  \vec\sigma_e(\vec\chi,\vec{\bar\sigma}) = \vec E^0 (\vec I-\vec B_e\vec X_e)\, \vec{\bar\epsilon}(\vec{\bar\sigma}).
\end{equation}
The microscopic initial stress tensor $\vec\sigma_e$ describes how the macroscopic strain $\vec{\bar\epsilon}$ is distributed in unit cell $Y$.
We note that \cref{eq:stressdistribution} already takes stress amplification (see \cite{ferrer2021stress}) into account.
The macroscopic strain is given as solution of the macroscopic constitutive equation
\begin{equation}\label{eq:macroconsteq}
\vec{\bar\epsilon}(\vec{\bar\sigma}) = (\vec E^H)^{-1} \vec{\bar\sigma}
\end{equation}
and $\vec X_e = [\vec\chi_e^1\ \vec\chi_e^2\ \vec\chi_e^3]$ contains the solutions of \cref{eq:linelahom}.
The macroscopic stress $\vec{\bar\sigma}$ is obtained via \cref{eq:linEla,eq:macrostress}.
Parameterization of the lattice geometry and linearity of the buckling analysis allows us to precompute homogenized properties. Thus, we can avoid solving homogenization problems during optimization (\cf \cref{ssec:scaledecoupling}).

We want to emphasize that the buckling analysis in \cref{eq:macrobuckling} and \cref{eq:microbuckling} yields pure global and pure local modes, respectively. In particular, cross-modes spanning over both scales and defined as a solution to Eq. (40) in the work of \cite{neves2019symbolic}, are not detected.
\section{Microstructure parameterization and upscaling model}\label{sec:parameterization}%
In the following section, we present our method to integrate the microscopic buckling load factor into a macroscopic scale optimization problem. The explanation is done for an exemplary lattice structure; however, we would like to stress again that the method can be applied to arbitrary base cell topologies.
We first demonstrate an approach to obtain the homogenized load factors of the lattice.

\begin{figure}[t]
  \centering
  \newsavebox{\unitcellbox}
  \savebox{\unitcellbox}{\includegraphics[width=\columnwidth]{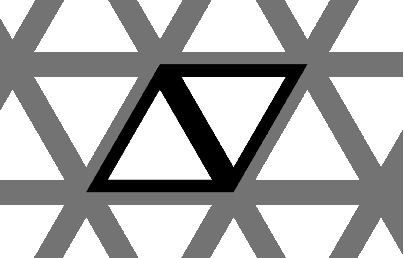}}
  \begin{tikzpicture}[>={Latex[length=5pt, width=3pt]}]
    \node[above right,inner sep=0pt] at (0,0) {\usebox{\unitcellbox}};
    \draw[|<->|] (0.213\wd\unitcellbox,0.22\ht\unitcellbox) -- +(0.365\wd\unitcellbox,0) node[pos=0.5, below] {$d$};
    \draw[->,red] (0.213\wd\unitcellbox,0.255\ht\unitcellbox) -- +(0.365\wd\unitcellbox,0) node[pos=0.5, above] {$R_1$};
    \draw[->,red] (0.213\wd\unitcellbox,0.255\ht\unitcellbox) -- +(0.1825\wd\unitcellbox,0.3161\wd\unitcellbox) node[pos=0.5, left] {$R_2$};
  \end{tikzpicture}
  \caption{Principal design of an equilateral triangular lattice structure with highlighted unit cell. The lattice geometry is given by primitive lattice vectors $R_1$ and $R_2$. For our enhanced version, see \cref{fig:repetitions}.}
  \label{fig:unitcell}
\end{figure}
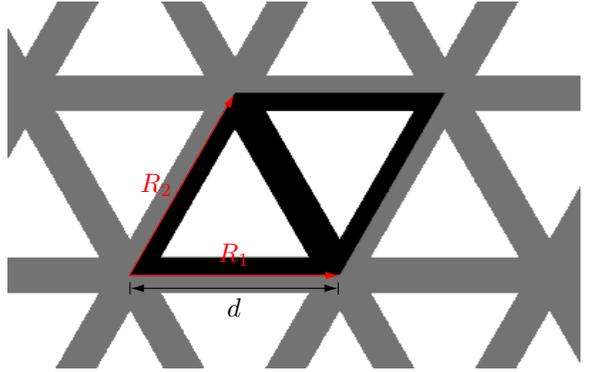

The exemplary lattice we chose consists of equilateral triangles with edge length $d$, as the principal design of our microstructure.
The primitive lattice vectors are thus given by $\vec{R_1}= (1,0)^\top$ and $\vec{R_2} = (1/2, \sqrt(3)/2)^\top$ (\cref{fig:unitcell}).
This design is known in literature to provide good macroscopic buckling strength \citep{clausen2016exploiting}.
To get a periodic unit cell, we take two of these triangles, which together form a parallelogram, including its shorter diagonal (see \cref{fig:unitcell}).
We parameterize the unit cell by one parameter $\rho$, which describes the relative volume or one minus porosity, respectively.
With this parameterization the homogenization procedures from \cref{sec:homogenization} can be written using two maps:
\begin{align}
E^H&:(0,1]\to\mathbb{S}^3,&&\rho\mapsto E^H(\rho),\label{eq:homlinela} \\
\lambda&:\RR^3\times(0,1]\to\RR,&&(\vec{\bar\sigma},\rho)\mapsto \lambda(\vec{\bar\sigma},\rho)\label{eq:hombuckling}
\end{align}
\cref{eq:homlinela} assigns each relative volume $\rho$ a symmetric, positive definite homogenized elasticity tensor computed from \cref{eq:homtensor},
while \cref{eq:hombuckling} maps a macroscopic stress and a relative volume to the resulting homogenized buckling load factor $\lambda(\vec{\bar\sigma},\rho)$ obtained by solving \cref{eq:microbuckling}.

As material parameters, we chose a Young's modulus of $10$\,Pa and a Poisson's ratio of $0.3$.

Next, we present three aspects that are important for the upscaling process:
the design of joints of lattice struts, the resolution of the finite element mesh on the microscopic level, and the number of unit cells in an RVE.
We note that the latter is automatically covered if a Floquet-Bloch approach along with a sufficiently fine discretization of the Brillouin zone is used. 

\subsection*{Rounded corners}

\begin{figure}
  \centering
  \begin{tikzpicture}[scale=6]
    \draw (0,0) -- (1.1,0);
    \draw (0,0) -- (0.5,{0.5*tan(60)});
    \draw[dotted] (0,0) -- (1.1,{1.1*tan(30)});

    \coordinate [circle,fill,inner sep=0.15em,label=above:{$\quad M_1$}] (m1) at (0.4,{0.4*tan(30)});
    \draw[blue] (m1) ++(150:.05) arc (150:270:.05);
    \draw[blue] (m1) ++(270:.05) coordinate(g11) -- +(0.7,0) coordinate(g12);
    \draw[blue] (m1) ++(150:.05) coordinate(g21) -- +(0.35,{0.35*tan(60)}) coordinate(g22);
    \draw (m1) -- ($(g11)!(m1)!(g12)$) node[pos=0.5,right]{$r_1$};
    \draw (m1) -- ($(g21)!(m1)!(g22)$);
    
    \coordinate [circle,fill,inner sep=0.15em,label=above:$\quad M_2$] (m2) at (0.95,{0.95*tan(30)});
    \draw[red,dashed] (m2) ++(150:.4) arc (150:270:.4);
    \draw[red,dashed] (m2) ++(270:.4) coordinate(h11) -- +(0.15,0) coordinate(h12);
    \draw[red,dashed] (m2) ++(150:.4) coordinate(h21) -- +(0.07,{0.07*tan(60)}) coordinate(h22);
    \draw[dashed] (m2) -- ($(h11)!(m2)!(h12)$) node[pos=0.5,right]{$r_2$};
    \draw[dashed] (m2) -- ($(h21)!(m2)!(h22)$) node[pos=0.5,above]{$r_2$};
    
%    \draw[-{Latex[length=5pt, width=3pt]}] (m1) ++(0.1,{0.1*tan(30)}) -- +(0.35,{0.35*tan(30)});

    \draw (0.6,0) -- ($(g11)!(0.6,0)!(g12)$) node[pos=0.5,right]{$w_1$};
    \draw (0.3,{0.3*tan(60)}) coordinate(l1) -- ($(g21)!(l1)!(g22)$) node[pos=0.5,above]{$w_1$};
    \draw[dashed] (1,0) -- ($(h11)!(1,0)!(h12)$) node[pos=0.5,right]{$w_2$};
    \draw[dashed] (0.45,{0.45*tan(60)}) coordinate(l2) -- ($(h21)!(l2)!(h22)$) node[pos=0.5,above]{$w_2$};
  \end{tikzpicture}
  \caption{To prevent stress concentrations, lattice corners are rounded by circular arcs, which are defined by a given radius. The figure shows the resulting lattice boundary for two configurations with different radii $r_1$ and $r_2$. The arc midpoint $M$ and lattice strut width $w$ are determined by the radius via geometric calculations, forcing the relative volume fraction of the lattice constant.}
  \label{fig:cornerrounding}
\end{figure}
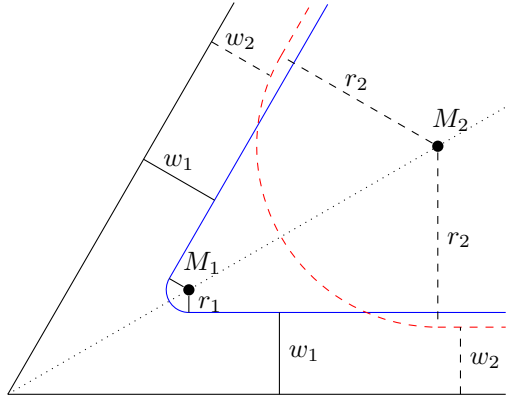

A unit cell design with sharp corners where the lattice struts meet leads to local stress concentrations at those corners.
To circumvent this issue, we round the corners with circular arcs with radius $r$, while keeping the overall volume of the structure constant (\cref{fig:cornerrounding}).
A lattice with rounded unit cells can be seen in \cref{fig:repetitions}.

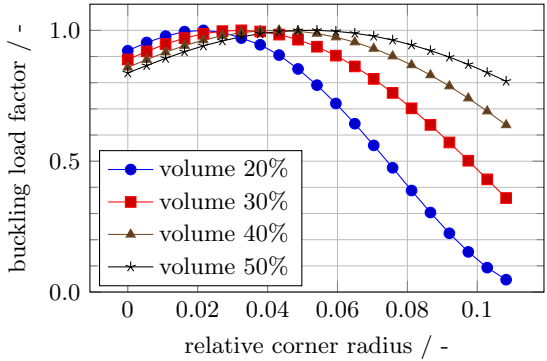
\begin{figure}[t]
  \centering
  \begin{tikzpicture}
    \pgfplotsset{
      width=\columnwidth,
      height=.25\textheight,
      xticklabel style={/pgf/number format/.cd,fixed,precision=2},
      yticklabel style={/pgf/number format/.cd,fixed,fixed zerofill,precision=1},
      legend style={at={(0.02,0.02)},anchor=south west}}
    \begin{axis}[
      ymin=0,
      xlabel=relative corner radius / -,
      ylabel=buckling load factor / -,
      minor y tick num=4,
      grid=both,
      cycle multiindex list={color\nextlist mark list}]
      \addplot table [x expr=\thisrowno{14}/32*sqrt(3), header=false, y expr=\thisrowno{0}/0.00697574326695788] {bending_study_v0.2.txt};
      \addlegendentry{volume $20\%$};
      \addplot table [x expr=\thisrowno{14}/32*sqrt(3), header=false, y expr=\thisrowno{0}/0.0274279850171493] {bending_study_v0.3.txt};
      \addlegendentry{volume $30\%$};
      \addplot table [x expr=\thisrowno{14}/32*sqrt(3), header=false, y expr=\thisrowno{0}/0.0756263851361581] {bending_study_v0.4.txt};
      \addlegendentry{volume $40\%$};
      \addplot table [x expr=\thisrowno{14}/32*sqrt(3), header=false, y expr=\thisrowno{0}/0.17047067816156] {bending_study_v0.5.txt};
      \addlegendentry{volume $50\%$};
    \end{axis}
  \end{tikzpicture}
  \caption{Normalized homogenized buckling load factor for $\vec{\bar\sigma})=(-1,0,0)$ and different relative volumes as a function of the radius of rounded corners. The radius is given relative to a unit cells edge length $d$.}
  \label{fig:radiusstudyLF}
\end{figure}

\begin{figure}[t]
  \centering
  \begin{tikzpicture}
    \pgfplotsset{
      width=\columnwidth,
      height=.25\textheight,
      xticklabel style={/pgf/number format/.cd,fixed,precision=3},
      yticklabel style={/pgf/number format/.cd,fixed,fixed zerofill,precision=1},
      legend style={at={(0.02,0.02)},anchor=south west}}
    \begin{axis}[
      ymin=0,
      xlabel=relative corner radius / -,
      ylabel=Young's modulus / Pa,
      minor y tick num=4,
      grid=both,
      cycle multiindex list={color\nextlist mark list}]
      \addplot table [x expr=\thisrowno{14}/32*sqrt(3), header=false, y expr=\thisrowno{3}/0.74488] {bending_study_v0.2.txt};
      \addlegendentry{volume $20\%$};
      \addplot table [x expr=\thisrowno{14}/32*sqrt(3), header=false, y expr=\thisrowno{3}/1.1952] {bending_study_v0.3.txt};
      \addlegendentry{volume $30\%$};
      \addplot table [x expr=\thisrowno{14}/32*sqrt(3), header=false, y expr=\thisrowno{3}/1.7263] {bending_study_v0.4.txt};
      \addlegendentry{volume $40\%$};
      \addplot table [x expr=\thisrowno{14}/32*sqrt(3), header=false, y expr=\thisrowno{3}/2.3769] {bending_study_v0.5.txt};
      \addlegendentry{volume $50\%$};
    \end{axis}
  \end{tikzpicture}
  \caption{Normalized homogenized Young's modulus for different relative volumes as a function of the radius of rounded corners. The radius is given relative to a unit cells edge length $d$.}
  \label{fig:radiusstudyE}
\end{figure}
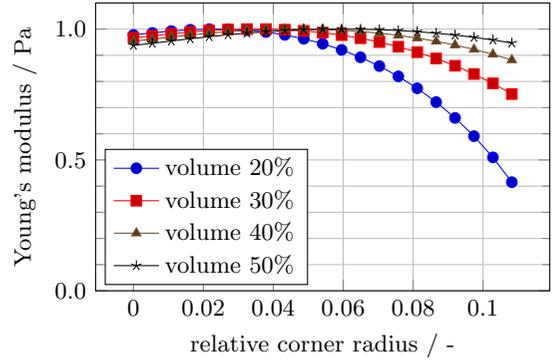

\begin{figure}[t]
  \centering
  \begin{tikzpicture}
    \pgfplotsset{
      width=\columnwidth,
      height=.25\textheight,
      xticklabel style={/pgf/number format/.cd,fixed,precision=2},
      yticklabel style={/pgf/number format/.cd,fixed,fixed zerofill,precision=2},
      legend style={at={(0.02,0.02)},anchor=south west}}
    \begin{axis}[
      ymin=0.8,
      xlabel=relative corner radius / -,
      ylabel=Poisson's ratio / -,
      grid=major,
      cycle multiindex list={color\nextlist mark list}]
      \addplot table [x expr=\thisrowno{14}/32*sqrt(3), header=false, y expr=\thisrowno{4}/0.35254] {bending_study_v0.2.txt};
      \addlegendentry{volume $20\%$};
      \addplot table [x expr=\thisrowno{14}/32*sqrt(3), header=false, y expr=\thisrowno{4}/0.35904] {bending_study_v0.3.txt};
      \addlegendentry{volume $30\%$};
      \addplot table [x expr=\thisrowno{14}/32*sqrt(3), header=false, y expr=\thisrowno{4}/0.36091] {bending_study_v0.4.txt};
      \addlegendentry{volume $40\%$};
      \addplot table [x expr=\thisrowno{14}/32*sqrt(3), header=false, y expr=\thisrowno{4}/0.35658] {bending_study_v0.5.txt};
      \addlegendentry{volume $50\%$};
    \end{axis}
  \end{tikzpicture}
  \caption{Normalized homogenized Poisson's ratio for different relative volumes as a function of the radius of rounded corners. The radius is given relative to a unit cells edge length $d$.}
  \label{fig:radiusstudyV}
\end{figure}
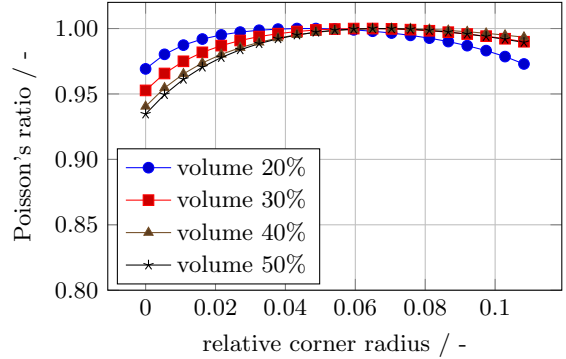

It turns out that the choice of the radius parameterizing the arcs has only minor effect on the Poisson's ratio (\cref{fig:radiusstudyV}).
The Young's modulus is affected mainly for small relative lattice volume (\cref{fig:radiusstudyE});
for a small volume, a large radius leads to very thin lattice struts due to volume preservation (\cf \cref{fig:cornerrounding}) and the lattice loses stiffness.
The impact on the homogenized buckling load factor is more significant as can be seen in \cref{fig:radiusstudyLF}.
For small radii, the buckling resistance increases with increasing radius, because stress concentrations are avoided, lattice struts get better supported, and their relative length to width ratio gets smaller.
For larger radii, the struts get thinner, their length to width ratio increases, and the load factor decreases.
For different volumes, the maximum of the smallest buckling load factor is achieved at different radii.
The data in \cref{fig:radiusstudyLF} suggests that rounding corners with a radius depending on the volume leads to good buckling resistance.
We note that we only investigated an uniaxial loading with $\vec{\bar\sigma})=(-1,0,0)$, and the optimal radius might be different for other stress situations.
To keep things simple, we thus decided to use a constant radius of $r=0.05d$ for our subsequent calculations. 

\subsection*{Finite element convergence on the microscale}
We discretize the microstructure by triangles with second order shape functions.
A convergence study with respect to the number of finite elements for homogenization, i.e., on the microscopic scale, can be seen in \cref{fig:feconvhomEv,fig:feconvhomblf}.
Observing that the convergence graphs are already very flat when approaching one million elements and taking into account that we want to keep the finite element error in the preprocessing small, we opt to choose for all subsequent homogenization procedures a discretization, which results for a density of $\rho=0.5$ in approximately one million finite elements per RVE. Geometries corresponding to lower densities are resolved using slightly less elements. 
Given this number of finite elements, a single analysis is usually completed within a few minutes on a standard work station.

\pgfplotstableread[header=false]{feconvhom.txt}{\feconvhom}
\begin{figure}[t]
  \centering
  \begin{tikzpicture}
    \pgfplotsset{
      width=0.58\columnwidth,
      set layers,
      scale only axis,
      xmin=0,
      x filter/.code={\pgfmathparse{#1/1000000}\pgfmathresult},
      legend style={at={(0.98,0.98)},anchor=north east}}
    \begin{axis}[
      axis y line*=left,
      ytick distance=0.0003,
      yticklabel style={/pgf/number format/.cd,fixed,fixed zerofill,precision=4},
      xlabel=number of finite elements / $10^6$,
      ylabel=Young's modulus / Pa]
      \addplot+[blue, mark=+, only marks] table [x index=0, y index=4] {\feconvhom};
      \label{plot:one}
    \end{axis}
    
    \begin{axis}[
      axis x line=none,
      axis y line*=right,
      ytick distance=0.0003,
      yticklabel style={/pgf/number format/.cd,fixed,fixed zerofill,precision=4},
      ylabel=Poisson's ratio]
      \addlegendimage{/pgfplots/refstyle=plot:one}\addlegendentry{Young's modulus}
      \addplot+[red, mark=x, only marks] table [x index=0, y index=5] {\feconvhom};
      \addlegendentry{Poisson's ratio}
    \end{axis}
  \end{tikzpicture}
  \caption{Homogenized Young's modulus and Poisson's ratio for different finite element discretizations of an RVE with a relative volume of $30\%$.}
  \label{fig:feconvhomEv}
\end{figure}
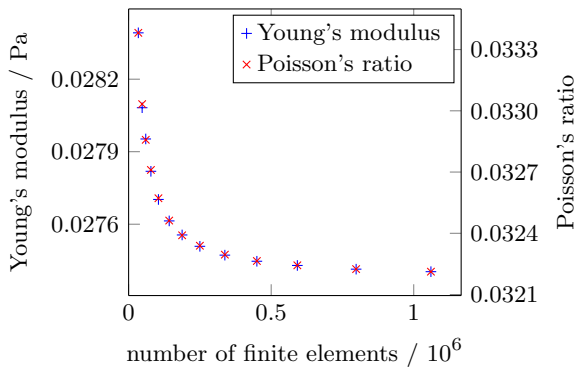
\begin{figure}[t]
  \centering
  \begin{tikzpicture}
    \pgfplotsset{
      width=\columnwidth,
      x filter/.code={\pgfmathparse{#1/1000000}\pgfmathresult}}
    \begin{axis}[
      yticklabel style={/pgf/number format/.cd,fixed,fixed zerofill,precision=4},
      xlabel=number of finite elements / $10^6$,
      ylabel=buckling load factor / -]
      \addplot+[mark=x,thick,only marks] table [x index=0, y index=3] {\feconvhom};
      \addlegendentry{buckling load factor};
    \end{axis}
  \end{tikzpicture}
  \caption{Homogenized buckling load factor for different finite element discretizations of an RVE with a relative volume of $30\%$.}
  \label{fig:feconvhomblf}
\end{figure}
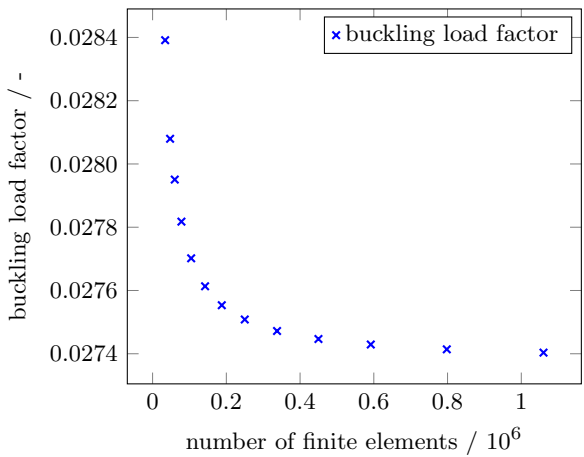

\subsection*{Choice of RVE size}
\begin{figure}[t]
  \centering
  \newsavebox{\repsbox}
  \savebox{\repsbox}{\includegraphics[width=\columnwidth]{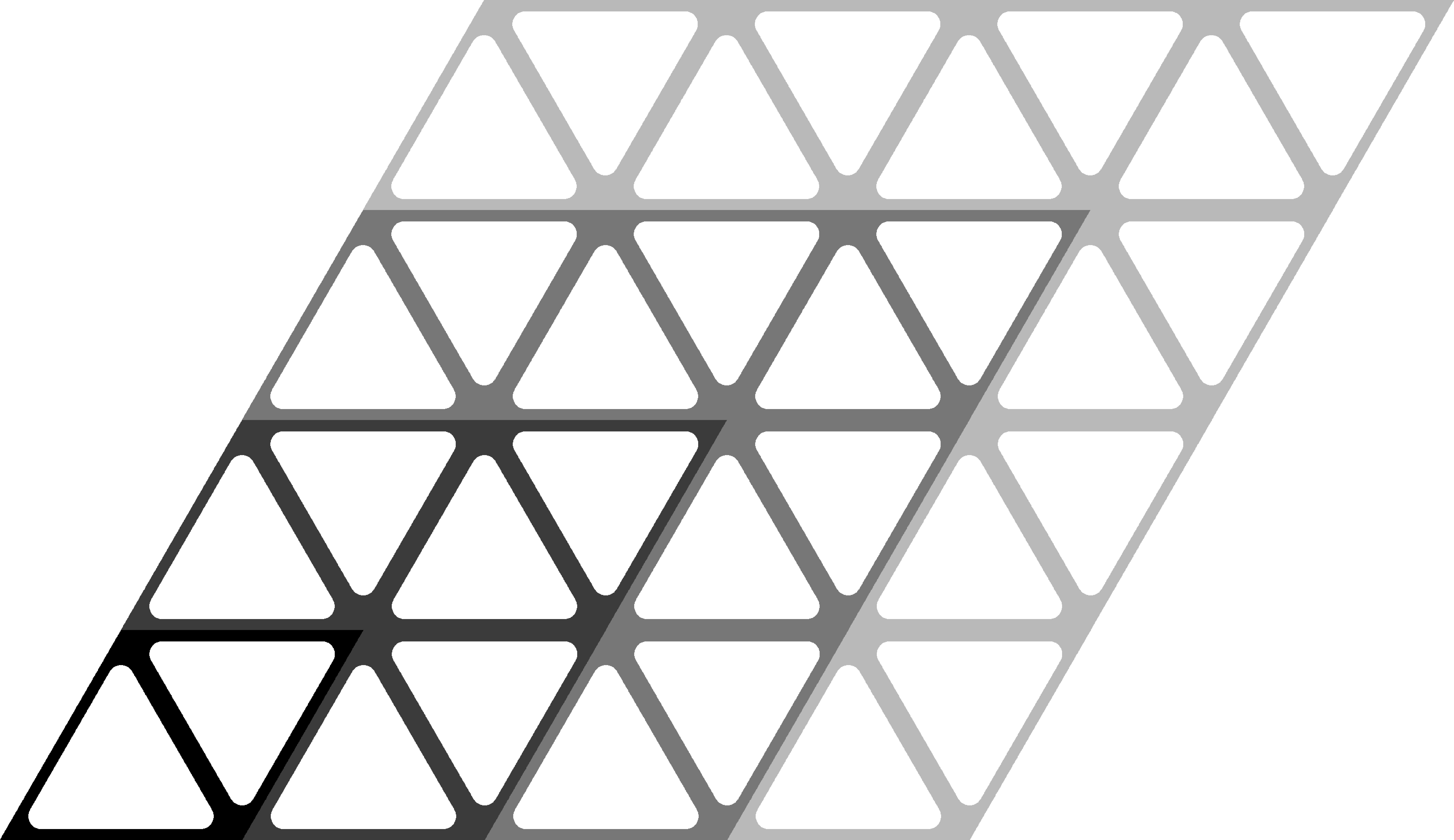}}
  \begin{tikzpicture}
    \node[above right,inner sep=0pt] at (0,0) {\usebox{\repsbox}};
    \draw[-{Latex[length=5pt, width=3pt]}] (0.8\wd\repsbox,0) -- +(0.15\wd\repsbox,0) node[below] {$x$};
    \draw[-{Latex[length=5pt, width=3pt]}] (0.8\wd\repsbox,0) -- +(0,0.15\wd\repsbox) node[right] {$y$};
  \end{tikzpicture}
  \caption{Different RVEs containing 1x1, 2x2, 3x3 and 4x4 unit cells and the coordinate system for homogenization.}
  \label{fig:repetitions}
\end{figure}
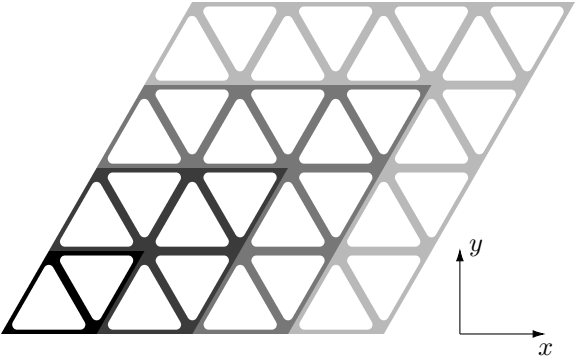

Asymptotic homogenization can only detect high-frequency modes, i.e., modes with a wave length that is smaller than the size of the representative volume element.
However, buckling modes might span over more than one cell.
To identify these modes also in a homogeneous lattice structure, we encompass more than one unit cell in the RVE. That is, the microstructure is considered $\tfrac{Y}{\kappa}$-periodic, where $\kappa^2$ is the number of unit cells inside the RVE.
RVEs containing various unit cells can be seen in \cref{fig:repetitions}.
It is noted that all modes of an RVE with a given number $\kappa$ of cell repetitions can also be found using an RVE comprised of $\ell \kappa \times \ell \kappa$ unit cells, where $\ell$ is an integer number, e.g., all modes of a $2\times2$ RVE appear also in the analysis of a $4\times4$ or $6\times6$ RVE.
Using an analogous argument, modes with periodicity $\tfrac{\kappa}{n_1}$ along $\vec{R_1}$ and periodicity $\tfrac{\kappa}{n_2}$ along $\vec{R_2}$ for $n_1, n_2\in\NN$ are captured using a $\kappa \times \kappa$ RVE.
Note that in practice there is a natural lower bound; this value is given by the resolution of the underlying finite element discretization.

We observe that especially an RVE with only one unit cell results in quite high homogenized load factors.
This is because the joint of the struts gets locked in rotation due to the periodicity of the RVE.
In an RVE with more than one cell and in the analysis of dehomogenized designs in \cref{ssec:dehomogenization}, joints of lattice structures are free to rotate (see \cref{fig:dehombucklingzoom}).

As we are interested in the smallest, i.e., critical, load factor of the lattice, we pick the minimal load factor with respect to cell repetitions:
\begin{equation}\label{eq:worstcaserepetitions}
\lambda(\vec{\bar\sigma},\rho) = \min_{\kappa\in\NN} \lambda_\kappa(\vec{\bar\sigma},\rho)
\end{equation}
In practice, one is not able to perform homogenization with $\kappa\in\NN$ cell repetitions, but has to resort to $\kappa\leq \text{K}$ with an appropriately chosen upper bound $\text{K}$ on the number of cell repetitions.
%Doing so, we observed that the minimum occurs for our specific unit cell at repetition count $\kappa=3$ for all investigated volume fractions.
%We note though, that this value might be different for other types of unit cells.
We would like to stress that for a given cell layout, the optimal $\kappa \in \{1,2,\ldots,K\}$ cannot, in general, be determined without testing essentially all choices and might even vary for fixed stress $\vec{\bar\sigma}$ but different volume fractions $\rho$.

\begin{figure}[t]
  \centering
  \includegraphics[trim=0 65 0 50,clip,width=.95\columnwidth]{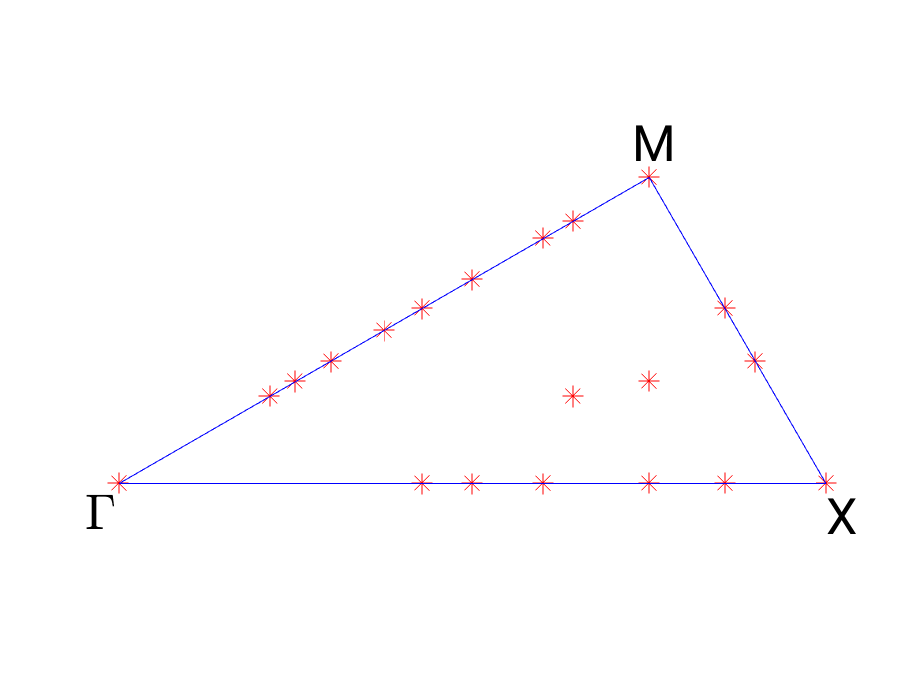}
  \caption{Wave vectors in the Brillouin zone (see \cite{thomsen2018buckling}), which correspond to homogenization with 1 to 7 cell repetitions in an RVE. Only 7 cell problems have to be solved to capture all modes corresponding to the marked wave vectors.}
  \label{fig:brillouin}
\end{figure}

We want to remark that the presented method of repeating the unit cell within the RVE is analogous to applying Floquet-Bloch theory with a special discretization of the Brillouin zone (see \cref{fig:brillouin}).
In Floquet-Bloch theory, periodicity conditions for modes $\vec\phi$ are given as
\begin{equation}\label{eq:FB_BC}
\vec\phi(\vec{y}+\vec{R_j}) = \vec\phi(\vec{y}) e^{i\vec{k}^\top \vec{R_j}}, \qquad j=1,2,
\end{equation}
where $i$ refers to the complex unity, $\vec{k}$ is the wave vector, and $\vec{y}$ is a location inside a lattice unit cell \citep{thomsen2018buckling}. 
Rather than solving cell problems with boundary conditions \cref{eq:FB_BC} for all $\vec{k} \in \RR^2$, in practice a number of $\vec{k}$ vectors on the boundary of the so called irreducible Brillouin zone (IBZ) is selected.
This corresponds to a discretization of the IBZ. 

Now, a $(\tfrac{\kappa}{n_1},\tfrac{\kappa}{n_2})$-periodic mode with $n_1, n_2\leq\kappa$ corresponds to a wave vector, which solves the system
\begin{equation}
\begin{pmatrix} \vec{R_1}^\top \\ \vec{R_2}^\top \end{pmatrix} \vec{k} = \tfrac{2\pi}{\kappa}\begin{pmatrix}n_1\\ n_2\end{pmatrix}.
\end{equation}
With that, we can plot all wave vectors corresponding to modes, covered by our $\kappa \times \kappa$ RVEs with $\kappa$ from 1 to $K$, into a Floquet-Bloch diagram with an outline of the IBZ. Doing so for $K=7$, we obtain the result depicted in \cref{fig:brillouin}.  

We emphasize that the worst-case model presented in the next \cref{ssec:worstcasemodel} is independent of the method that is used to obtain the homogenized buckling load factor.

\subsection{Worst-case model}\label{ssec:worstcasemodel}
This subsection describes our novel method to integrate the microscopic buckling load factor on the macroscopic scale.
We stress that it is valid not only for our exemplary lattice but for any arbitrary, parameterized microstructure.

Due to linearized buckling analysis (\cref{eq:microbuckling,eq:stressmatrixmicro,eq:stressdistribution,eq:macroconsteq}) the homogenized load factor depends linearly on the macroscopic stress:
\begin{equation}\label{eq:stressnorm}
\lambda(\vec{\bar\sigma},\rho) = \tfrac{1}{\|\vec{\bar\sigma}\|} \min_{\kappa\in\NN} \lambda_\kappa\left(\tfrac{\vec{\bar\sigma}}{\|\vec{\bar\sigma}\|},\rho\right)
\end{equation}
That is, homogenization can be conducted with macroscopic unit stress (in some given norm) and later, the homogenized load factor has to be divided by $\|\vec{\bar\sigma}\|$.
It is thus sufficient to examine stresses on the unit sphere surface $S^2$ instead of the whole three-dimensional stress space $(\sigma_{xx},\sigma_{yy},\sigma_{xy})$.
We define
\begin{align}\label{eq:fullmodel}
\wbar\lambda:\ &S^2\times(0,1] \to\RR,\notag\\
&\wbar\lambda(\vec{\bar\sigma},\rho) = \min_{\kappa\in\NN} \lambda_\kappa(\vec{\bar\sigma},\rho),
\end{align}
which assigns a homogenized buckling load factor to each unit stress combined with a relative volume.

\begin{figure}[t]
  \centering
  \includegraphics[width=.98\columnwidth]{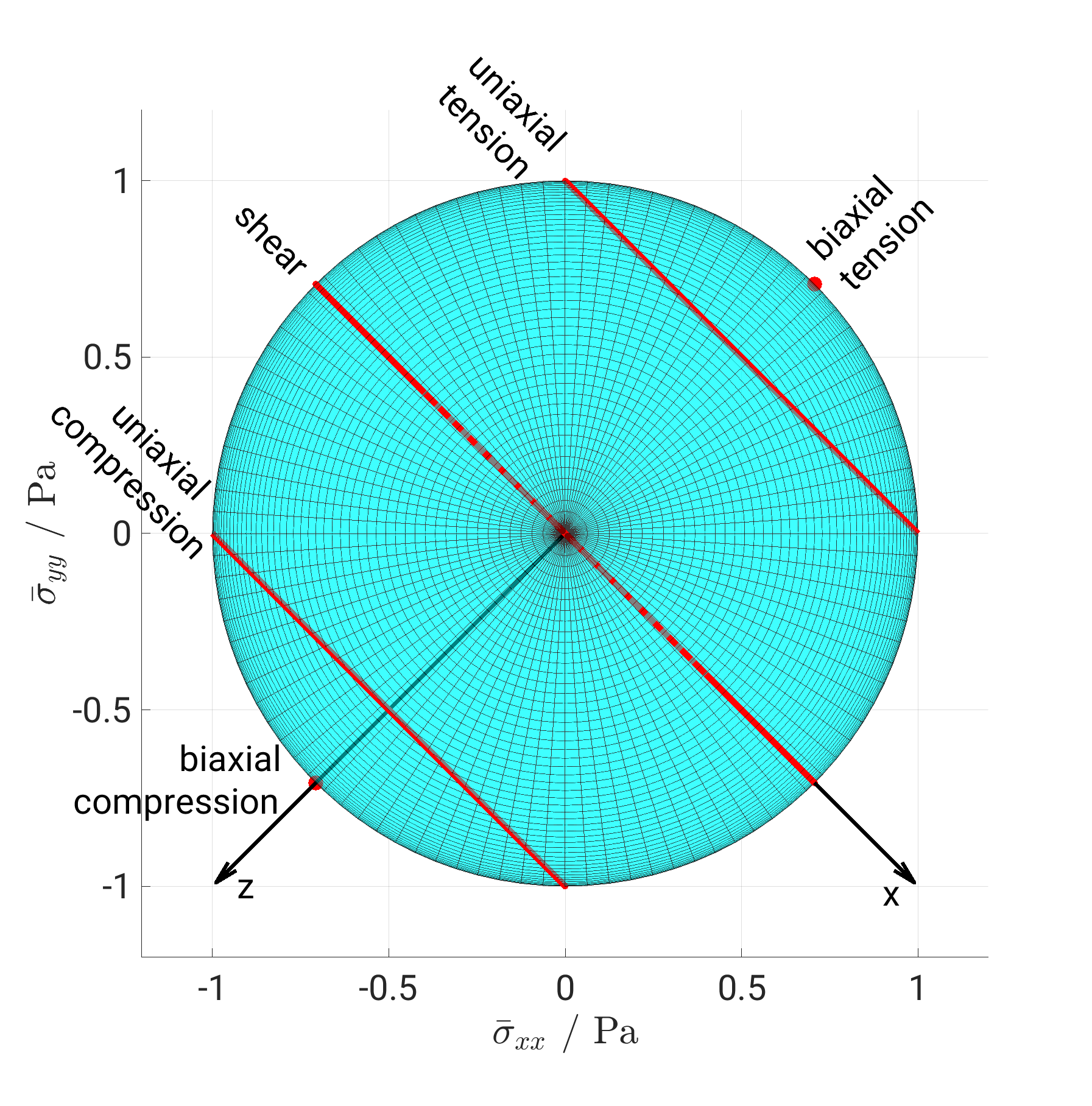}
  \caption{Representation of unit stresses as a unit sphere surface ($\vec{\bar\sigma})_{xy}$ out of plane axis). Special stress types form circles (uniaxial/shear stress) or single points (biaxial stress) on the surface. Unit stresses are parameterized by spherical coordinates with zenith reference~$z$ and azimuthal reference~$x$.}
  \label{fig:stresssphere}
\end{figure}
We use a spherical coordinate system to characterize unit stresses on $S^2$:
the zenith reference (z-axis) is the axis from the origin through the biaxial compression stress $\vec\sigma=(-1,-1,0)$ and the azimuth reference (x-axis) is the axis from the origin through $\vec\sigma=(1,-1,0)$ (\cref{fig:stresssphere}).
In this coordinate system, biaxial compression and tension stress conform to north and south pole, while all pure shear stresses rest on the equator.
The inclination (or latitude if thinking of geographical coordinates) characterizes the type of stress:
the rotation invariant biaxial compression and tension stresses conform to poles, and other special types, e.g., uniaxial and shear stresses form circles on the unit sphere surface (circle of latitude).
The azimuthal angle (longitude) describes the rotation of the applied macroscopic stress $\vec{\bar\sigma}$ relative to the RVE. 
For an additional explanation please refer to the video in Online Resource 1.
%We note that the worst-case approach works for abritrary parametrizations of the unit sphere, however, our choice has a direct interpretation in the coordinate system used for homogenization.

Common literature reduces the stress space even further.
Under the assumption of isotropic buckling behavior, only biaxial loading without shearing component but varying the $\tfrac{\sigma_{xx}}{\sigma_{yy}}$-ratio is investigated, or uniaxial loading with all possible load orientations relative to the investigated microstructure is applied \citep{bluhm2020nonlinear}.
In contrast, we compute the buckling yield surface for the whole unit stress surface $S^2$.
Thus, our method is applicable to arbitrary, parameterized microstructures with isotropic or anisotropic buckling properties.

The homogenized load factors for uniaxial compression, biaxial compression and shear stress for an RVE with a relative volume of $30\%$ and three cell repetitions are shown in \cref{fig:homogenizedlf}.
We can clearly see that the symmetry of the unit lattice cell is reflected in the load factors.
The shape of the uniaxial loading case matches very well with results by \cite{bluhm2020nonlinear}.

\pgfplotstableread[header=false]{uniaxial_comp_v0.3.txt}{\uniaxialc}
\pgfplotstableread[header=false]{uniaxial_tens_v0.3.txt}{\uniaxialt}
\pgfplotstableread[header=false]{shear_v0.3.txt}{\shear}
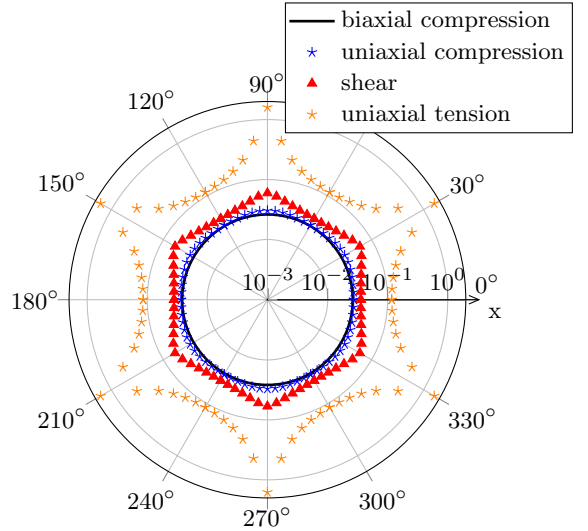
\begin{figure}[t]
  \begin{tikzpicture}
    \pgfplotsset{width=.9\columnwidth}
    \begin{polaraxis}[
      xticklabel={$\pgfmathprintnumber\tick^\circ$},
      xticklabel style={yshift={(\ticknum==1)*.6em}},
      ymin=1e-3, ymax=2,
      ytick={0.001,0.01,0.1,1},
      yticklabels={$10^{-3}$,$10^{-2}$,$10^{-1}$,$10^0$},
      y coord trafo/.code=\pgfmathparse{log10(#1)+3},
      y coord inv trafo/.code=\pgfmathparse{log10(#1)-3},
      yticklabel style={/pgf/number format/.cd,fixed,fixed zerofill,precision=2},
      extra x ticks={0},
      extra x tick labels={x},
      extra x tick style={tick label style={yshift=-.5em,xshift=.5em}},
      xminorgrids,
      clip=false,
      legend style={inner xsep=1pt, inner ysep=1pt, at={(1.27,1.25)}, anchor=north east}]
      \addplot[black,domain=-180:180,samples=90,line width=1pt] (x,0.0263078985021975);
      \addplot[blue,mark=star,legend image post style={only marks}] table [x expr=\thisrowno{14}*5, y index=0, header=false, only marks] {\uniaxialc};
      \addplot[blue,mark=star] table [x expr=\thisrowno{14}*5+ 60, y index=0, header=false, only marks] {\uniaxialc};
      \addplot[blue,mark=star] table [x expr=\thisrowno{14}*5+120, y index=0, header=false, only marks] {\uniaxialc};
      \addplot[blue,mark=star] table [x expr=\thisrowno{14}*5+180, y index=0, header=false, only marks] {\uniaxialc};
      \addplot[blue,mark=star] table [x expr=\thisrowno{14}*5+240, y index=0, header=false, only marks] {\uniaxialc};
      \addplot[blue,mark=star] table [x expr=\thisrowno{14}*5+300, y index=0, header=false, only marks] {\uniaxialc};
      \addplot[red,mark=triangle*,legend image post style={only marks}] table [x expr=\thisrowno{14}*5, y index=0, header=false, only marks] {\shear};
      \addplot[red,mark=triangle*] table [x expr=\thisrowno{14}*5+ 60, y index=0, header=false, only marks] {\shear};
      \addplot[red,mark=triangle*] table [x expr=\thisrowno{14}*5+120, y index=0, header=false, only marks] {\shear};
      \addplot[red,mark=triangle*] table [x expr=\thisrowno{14}*5+180, y index=0, header=false, only marks] {\shear};
      \addplot[red,mark=triangle*] table [x expr=\thisrowno{14}*5+240, y index=0, header=false, only marks] {\shear};
      \addplot[red,mark=triangle*] table [x expr=\thisrowno{14}*5+300, y index=0, header=false, only marks] {\shear};
      \addplot[orange,mark=star,legend image post style={only marks}] table [x expr=\thisrowno{14}*5, y index=0, header=false, only marks] {\uniaxialt};
      \addplot[orange,mark=star] table [x expr=\thisrowno{14}*5+ 60, y index=0, header=false, only marks] {\uniaxialt};
      \addplot[orange,mark=star] table [x expr=\thisrowno{14}*5+120, y index=0, header=false, only marks] {\uniaxialt};
      \addplot[orange,mark=star] table [x expr=\thisrowno{14}*5+180, y index=0, header=false, only marks] {\uniaxialt};
      \addplot[orange,mark=star] table [x expr=\thisrowno{14}*5+240, y index=0, header=false, only marks] {\uniaxialt};
      \addplot[orange,mark=star] table [x expr=\thisrowno{14}*5+300, y index=0, header=false, only marks] {\uniaxialt};
      \legend{biaxial compression,uniaxial compression,,,,,,shear,,,,,,uniaxial tension,,,,,,};
      \node (origin) at (axis cs:0,1e-3){};
      \node (xaxis) at (axis cs:0,5.0){};
      \draw[-angle 45] (origin) -- (xaxis);
    \end{polaraxis}
  \end{tikzpicture}
  \hfill
  \caption{Homogenized buckling load factors for an RVE with 3x3 unit cell repetitions and a relative volume of $30\%$. Note the logarithmic scaling of the radial axis. The displayed angle is the azimuthal angle from the spherical reference coordinate system (see \cref{fig:stresssphere}) and corresponds to the rotation of the stress around the RVE. The symmetry of the unit cell is reflected in the load factors.}
  \label{fig:homogenizedlf}
\end{figure}

Next, we coalesce all data in a worst-case model.
Having evaluated homogenized load factors for different cell repetitions $\kappa\in\NN$ and all stresses on the unit stress sphere $\vec{\bar\sigma}\in S^2$, i.e., for all stress types and directions,
we select the smallest buckling load factor with respect to all unit stresses and number of cell repetitions:
\begin{align}\label{eq:worstcase}
\lambda_{wc}:\ &(0,1]\to\RR,\notag\\
&\lambda_{wc}(\rho)=\min_{\substack{\vec{\bar\sigma}\in S^2,\\ \kappa\in\NN}} \lambda_\kappa(\vec{\bar\sigma},\rho)
\end{align}
This worst-case model depends only on the local volume fraction and no longer on the local stress type or direction.
We note that in practice the unit sphere is discretized and the number of cell repetitions bounded from above.
The worst case is thus only a worst case with respect to the discretization resolution and maximal cell repetitions.
We found, however, that the homogenized buckling load factors for our exemplary lattice cell have high regularity with respect to the stress variable.

If homogenized load factors were obtained via Floquet-Bloch theory \citep{neves2002topologycriteria} instead of using different cell repetitions, \cref{eq:worstcase} would contain a minimization over all possible wave vectors~$k$ in the Brillouin zone~$\mathcal{B}$ in lieu of cell repetitions:
\begin{equation}
\lambda_{wc}(\rho)=\min_{\substack{\vec{\bar\sigma}\in S^2,\\ \vec{k}\in\mathcal{B}}} \wbar\lambda_{\vec{k}}(\vec{\bar\sigma},\rho)
\end{equation}

%We observed for the exemplary equilateral triangular lattice, that the minimal homogenized load factor for each investigated volume fraction occurs for a biaxial compression stress state, i.e., compression is more critical than tension.
%However, this might be different for other lattices.

\subsection{Decoupling of micro- and macroscale}\label{ssec:scaledecoupling}
The parameterization of the unit cell allows for effective decoupling of the micro- and macroscopic scales \citep{bendsoe1988generating}.
That is, we discretize the parameter space $(0,1]$ for the relative volume and precompute homogenized properties (elasticity tensor from \cref{eq:homtensor} and microscopic buckling load factor from \cref{eq:microbuckling}) for the resulting parameter set (see also \cref{sec:numericalexamples}).
For optimization of structures on the macroscopic scale, we then apply an interpolation model to these precomputed properties.
In other words, we replace the maps $E^H$ \cref{eq:homlinela} and $\lambda$ \cref{eq:hombuckling} by
\begin{align}
E^I&:(0,1]\to\mathbb{S}^3,&&\rho\mapsto E^I(\rho),\label{eq:intlinela} \\
\lambda^I&:\RR^3\times(0,1]\to\RR,&&(\vec{\bar\sigma},\rho)\mapsto \lambda^I(\vec{\bar\sigma},\rho),
\end{align}
where $E^I$ and $\lambda^I$ are approximations of $E^H$ and $\lambda$, respectively.

To construct $\lambda^I$, we interpolate the worst-case load factors obtained from \cref{eq:worstcase} with respect to the density variable:
\begin{equation}\label{eq:intbucklingwc}
\lambda_{wc}^I:(0,1]\to\RR, \quad\rho\mapsto\lambda_{wc}^I(\rho)
\end{equation}

We obtain the worst-case microscopic buckling load factor associated with macroscopic stress $\vec{\bar\sigma}$ by dividing by the norm of this stress (see \cref{eq:stressnorm}):
\begin{equation}\label{eq:intbuckling}
\lambda^I:\RR^3\times (0,1]\to\RR, \quad (\vec{\bar\sigma},\rho)\mapsto \tfrac{1}{\|\vec{\bar\sigma}\|}\lambda_{wc}^I(\rho)
\end{equation}

For gradient based optimization, we need a differentiable interpolation model.
Following \cite{bendsoe1988generating} we apply a piecewise interpolation strategy.
More precisely, we employ piecewise cubic Hermite interpolation \citep{birkhoff1968piecewise} for both, the homogenized elasticity tensor \cref{eq:intlinela} as well as the buckling load factor \cref{eq:intbucklingwc}.
This scheme yields a continuously differentiable approximation that is composed of uniquely defined cubic polynomials between the provided data points and is exact at these points.
However, other interpolation schemes using, e.g., tangent or RAMP (Rational Approximation of Material Properties, \cite{stolpe2001alternative}) functionals are possible with the downside of lower accuracy, especially for high porosity values.

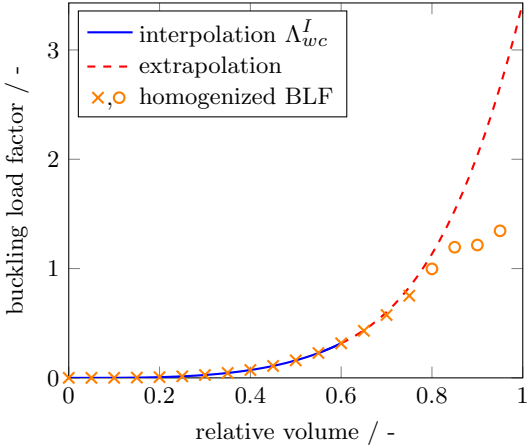
\begin{figure}[t]
  \pgfplotstableread[header=true]{worstCaseData.txt}{\worstCaseData}
  \pgfplotstablegetrowsof\worstCaseData \pgfmathsetmacro\numberofrows{\pgfplotsretval-9}

  \pgfmathdeclarefunction{mypoly}{7}{%
    % #1: a0, #2: a1, #3: a2, #4: a3, #5: x, #6: xl, #7: xu
    \pgfmathparse{#1+#2*(#5-#6)/(#7-#6)+#3*((#5-#6)/(#7-#6))^2+#4*((#5-#6)/(#7-#6))^3}%
  }%

  \begin{tikzpicture}
    \pgfplotsset{width=\columnwidth}
    \begin{axis}[
      domain=0:1,
      xlabel=relative volume / -,
      ylabel=buckling load factor / -,
      legend style={at={(0.02,0.98)},anchor=north west},
      enlargelimits=false,
      good blf/.style = {orange, mark=x, thick, mark size=3pt},
      bad blf/.style = {orange, mark=o, thick, mark size=2pt},
      combo legend/.style={
        legend image code/.code={
          \draw[/pgfplots/good blf] plot coordinates {(0.4em,0em)};
          \draw plot coordinates {(0.5em,-3pt)} node {,};
          \draw[/pgfplots/bad blf] plot coordinates {(1.2em,0em)};
        }
      }]
      \pgfplotsinvokeforeach{0,...,\numberofrows}{
        \pgfplotstablegetelem{#1}{[index]0}\of\worstCaseData \pgfmathsetmacro{\xZero}{\pgfplotsretval}
        \pgfplotstablegetelem{\the\numexpr #1+1}{[index]0}\of\worstCaseData \pgfmathsetmacro{\xOne}{\pgfplotsretval}
        \pgfplotstablegetelem{#1}{[index]4}\of\worstCaseData \pgfmathsetmacro{\aZero}{\pgfplotsretval}
        \pgfplotstablegetelem{#1}{[index]5}\of\worstCaseData \pgfmathsetmacro{\aOne}{\pgfplotsretval}
        \pgfplotstablegetelem{#1}{[index]6}\of\worstCaseData \pgfmathsetmacro{\aTwo}{\pgfplotsretval}
        \pgfplotstablegetelem{#1}{[index]7}\of\worstCaseData \pgfmathsetmacro{\aThree}{\pgfplotsretval}
        \addplot[domain=\xZero:\xOne,blue,thick,forget plot]{mypoly(\aZero,\aOne,\aTwo,\aThree,x,\xZero,\xOne)};
      }
      \addplot[domain=\xZero:\xOne,blue,thick]{mypoly(\aZero,\aOne,\aTwo,\aThree,x,\xZero,\xOne)};
      \addlegendentry{interpolation $\Lambda_{wc}^I$}
      \addplot[domain=\xOne:1,red,thick,dashed]{mypoly(\aZero,\aOne,\aTwo,\aThree,x,\xZero,\xOne)};
      \addlegendentry{extrapolation}
      \addplot[forget plot, good blf, legend image post style={only marks}, select coords between index={0}{15}] table [x=x, y=loadfactor, header=true, only marks] {\worstCaseData};
      \label{plot:wcdgood};
      \addplot[forget plot, bad blf, legend image post style={only marks}, select coords between index={16}{21}] table [x=x, y=loadfactor, header=true, only marks] {\worstCaseData};
      \label{plot:wcdbad};
      \addlegendimage{combo legend}\addlegendentry{homogenized BLF}
    \end{axis}
  \end{tikzpicture}
  \caption{Homogenized worst-case buckling load factors and the interpolating curve. For high relative volumes, numerical issues arise (marked by an o). Thus, only values up to $0.6$ are interpolated, and quadratic extrapolation is applied above.}
  \label{fig:worstcasemodel}
\end{figure}

Under the piecewise Hermite approach, $E^I$ is constructed by individual interpolation of each coefficient of the elasticity tensor. 
The first order derivatives of the homogenized elasticity tensor and homogenized buckling load factor with respect to $\rho$ are approximated by finite differences.
For the homogenized load factors, we get numerical issues for high relative volumes.
For these volumes, the unit cell is very resistant to buckling, but the iterative eigenvalue problem solver (ARPACK, version 3.7.0 \citep{lehoucq1998arpack}) yields artificial modes, which are caused by the buckling of individual finite elements.
Hence, we only interpolate data points below $60\%$ relative volume.
At the boundary of this sample space, second-order central finite differences are not available, and thus only quadratic polynomials are used in the two outer subintervals, i.e., $[0,0.05]$ and $[0.55,0.6]$.
For relative volumes above the chosen threshold of $60\%$ we extrapolate using the quadratic function obtained for the subinterval $[0.55,0.6]$.
It is noted, however, that in the optimized designs presented in \cref{sec:numericalexamples} microbuckling occurs only for significantly lower densities of $60\%$.
The interpolated worst-case model $\lambda_{wc}^I$ for the homogenized microscopic buckling load factor is shown in \cref{fig:worstcasemodel}.
The presented interpolation scheme leads to sufficiently differentiable functions to perform gradient-based optimization.
%Microstructures with high relative volume fractions are more likely to fail due to plastic yielding than local buckling \citep{timoshenko1989theory}.
%Thus, one might choose other extrapolation functions like in the work of \cite{christensen2022multiscale}.

The error of the approximated microscopic buckling load factor $\lambda^I$ comprises several individual errors:
the error introduced by finite element discretization on the microscopic level to solve the homogenization problems,
the error arising from restriction to the worst case, and
the error resulting from the interpolation.
The discretization and interpolation errors can easily be controlled by using finer finite element meshes on the microscopic scale and finer interpolation grids.
Thus, the worst-case error has the highest significance.
We will come back to this when we discuss numerical examples in \cref{ssec:worstcaseimpact}.
\begin{remark}\label{rem:wcfree} The worst-case error can be avoided if the worst-case model \cref{eq:worstcase} and univariate interpolation \cref{eq:intbucklingwc} are replaced by a discretization and trivariate interpolation of \cref{eq:fullmodel} in the combined stress and density space $S^2\times(0,1]$.

In gradient-based optimization context, a continuously differentiable interpolation scheme is essential.
Possible techniques to realize that for this worst-case free approach include piecewise cubic Hermite interpolation \citep{birkhoff1968piecewise} or interpolation on sparse grids based on cubic B-splines \citep{valentin2020gradient}.
When applying the worst-case free approach in a three-dimensional setting, due to the curse of dimensionality, the differentiable sparse grid approach is preferable, as stress has six independent entries there, which result in five parameters to represent unit stress, requiring interpolation to be carried out in a six dimensional space.
\end{remark}

\section{Sizing optimization}\label{sec:sizingopt}%
In this section, we formulate two-scale sizing optimization problems, which will be solved in \cref{sec:numericalexamples}.
In contrast to topology optimization, where usually a solid/void design is of interest, we vary the local lattice volume fraction between $10\%$ and $100\%$.
This corresponds to a scaling of the width of the lattice struts.

We want to achieve structures that are resistant to given loadings both with respect to their stiffness and their buckling strength.
To maximize buckling strength in our two-scale approach, we have to take into account buckling of the homogenized overall component as well as buckling of the microstructure.
This leads to a multi-objective optimization problem.
For a macroscopic domain, discretized by $\text{M}$ finite elements, it can be stated that in terms of mechanical compliance $c$, macroscopic load factors $\Lambda_\ell, \ell=1,\ldots,L$ and microscopic buckling load factors $\lambda_e^I, e=1,\ldots,\text{M}$:
\begin{align}\label{eq:multiobj}
\max_{\vec\rho\in\Uad} g\left(-c,\Lambda_1,\ldots,\Lambda_\text{L},\lambda_1^I,\ldots,\lambda_\text{M}^I\right)(\vec\rho).
\end{align}
The macroscopic load factors are given as solutions of \cref{eq:macrobuckling} and a prediction for the microscopic buckling load factor $\lambda_e^I$ for each element $e$ is obtained from the interpolated worst-case model \cref{eq:intbuckling}.
The design variable $\vec\rho$ represents the local volume fraction of the homogenized lattice.
Via the homogenization formulas, this can be mapped to mechanical properties of a (periodic) microstructure, e.g., elasticity tensor \cref{eq:homtensor} and buckling load factor \cref{eq:intbuckling}, which uses \cref{eq:microbuckling}.
$g$ is an aggregating function and $\Uad$ is the admissible set
\begin{equation}
    \Uad = \left\{ \vec\rho\in\RR^\text{M} : \rho_\text{min}\leq\rho_e\leq 1 , \sum_{e=1}^\text{M} \rho_e \leq V\right\}.
\end{equation}
We include the first few $\text{L}\ll\text{M}$ macroscopic load factors in the objective to handle potential mode switching and multiple eigenvalues.
Note that minimization of the compliance corresponds to maximization of its negative value.
We treat the multi-objective problem \cref{eq:multiobj} with the $\epsilon$-constraint method \citep{mavrotas2009effective}. In this, a point on the Pareto front is obtained by solving the following problem:
\begin{equation}
\begin{split}
\max_{\vec\rho\in\Uad}\ & \tilde g\left(\Lambda_1,\ldots,\Lambda_\text{L},\lambda_1^I,\ldots,\lambda_\text{M}^I\right)(\vec\rho), \\
\st\ & c(\vec\rho) = c_0,
\end{split}
\end{equation}
with an imposed compliance value $c_0$.

We want to maximize the smallest load factor to achieve a good buckling strength and thus choose $\tilde g$ to be the $\min$ function. To get rid of the non-smooth character of the latter, we apply a bound formulation of this problem, as suggested by \cite{bendsoe2003topology}, and end up with the following optimization problem:
\begin{align}\label{prob:macroopt}
\max_{s\in\RR,\vec\rho\in\Uad}\ s,\ & \\
\st\ \Lambda_\ell(\vec\rho) &\geq s, \qquad \ell=1,\ldots,\text{L} \label{eq:probmacrobucklingconstraint}\\
\lambda_e^I(\vec\rho) &\geq s, \qquad e=1,\ldots,\text{M} \label{eq:probmicrobucklingconstraint}\\
c(\vec\rho) &= c_0, \label{eq:probcomplianceconstraint}
\end{align}
With the bound formulation we experienced no problems with multiple eigenvalues or clustering of eigenvalues in our numerical experiments, if $L$ was chosen sufficiently large.
%For a more rigorous handling of multiple eigenvalues we refer to the suggestions in the book of \cite{bendsoe2003topology}.
Alternatively, other formulations like a smooth minimum via \cite{kreisselmeier1980systematic} could be considered.

In \cref{ssec:optimization}, we look at three variations of this problem:
\begin{enumerate}[label=\Alph*)]
  \item\label{itm:globBuck} We ignore \cref{eq:probmicrobucklingconstraint},
    i.e., we maximize buckling on the macroscopic scale with a compliance constraint but do not take buckling on the microscopic scale into account during optimization. 
  \item\label{itm:locBuck} We drop \cref{eq:probmacrobucklingconstraint},
    i.e., we only optimize microscopic buckling under a compliance constraint but disregard macroscopic buckling.
  \item\label{itm:globLocBuck} We investigate the problem as given,
    i.e., we maximize buckling on both scales with respect to a given compliance.
\end{enumerate}
We compute a representation of the Pareto optimal set with $51$ Pareto optimized solutions for each case.
For this, we first compute a solution to the pure compliance minimization problem
\begin{equation}\label{eq:complianceprob}
c^* = \argmin_{\vec\rho\in\Uad} c(\vec\rho).
\end{equation}
Then we solve \ref{itm:globBuck} - \ref{itm:globLocBuck} without the compliance constraint \cref{eq:probcomplianceconstraint} and evaluate the compliance values $c_A,c_B$ and $c_C$ of the resulting designs to get the extreme points on the Pareto set.
Afterwards, we discretize the intervals $[c^*,c_A],[c^*,c_B]$ and $[c^*,c_C]$ equidistantly, e.g., for \ref{itm:globBuck}
\begin{equation}
c_0^j = c^* + j\,(c_A-c^*)/50,\quad j=0,\ldots,50.
\end{equation}
The representation of the Pareto optimal set is then given by solutions to \ref{itm:globBuck} - \ref{itm:globLocBuck} with compliance constraint $c(\vec\rho)=c_0^j$ in \cref{eq:probcomplianceconstraint}.

\section{Numerical examples}\label{sec:numericalexamples}%
In this section we present the results of numerical experiments for the two-scale sizing problems stated in \cref{sec:sizingopt}.
We first conduct a pre-study to develop a better understanding of later optimized designs.
After that, solutions to the multi-objective optimization problems will be investigated.

To construct the worst-case model, we have to conduct buckling homogenization with applied macroscopic stress.
For this, we discretize the unit stress sphere in a spherical coordinate system (see \cref{ssec:worstcasemodel}).
We discretize the azimuthal angle, which describes the rotation of the applied stress around the RVE, with $5^\circ$ steps.
Exploiting the symmetry of our unit cell, it suffices to investigate the interval from $0^\circ$ to $30^\circ$ degrees. 
For the inclination, which describes the type of stress, we choose $11.25^\circ$ steps in the interval $[0^\circ,180^\circ)$.
We discretize the relative volume with $0.05$ steps and solve homogenization problems for four, five, six, and seven cell repetitions.
Modes obtained with an RVE consisting of two and three unit cells can also be detected in an RVE with six unit cell repetitions. Likewise, the modes for an 1x1 RVE are covered by all other RVEs (see \cref{sec:parameterization}).
This leads to $448$ cell problems for each chosen relative volume and $9408$ in total.
Note that these simulations are independent of each other and can be run in parallel.
We discretize the microscopic domain by approximately one million finite elements per RVE.
As already mentioned in \cref{sec:parameterization}, the material properties are given by a Young's modulus of $10$\,Pa and a Poisson's ratio of $0.3$.
It is noted that the motivation for the choice of the Young's modulus was an improved numerical behaviour, in particular when solving the associated eigenvalue problem on the microscopic scale.
Conversely, the results are invariant with respect to this choice.

The eigenvalue problems are solved by ARPACK, version 3.7.0 \citep{lehoucq1998arpack}.
%The minimal homogenized buckling load factor was found at $\kappa=3$ cell repetitions in the RVE and biaxial stress for each volume (\cf \cref{eq:worstcaserepetitions}).
%Note that this value strongly depends on the chosen unit cell and might differ for other lattices.
To solve optimization problems, we apply SNOPT, version 7.2.8 \citep{gill2005snopt}, which employs a sequential quadratic programming method.

\newsavebox{\simpcbbox}
\savebox{\simpcbbox}{\includegraphics[trim={350px 75px 0px 75px},clip,height=.4\textheight]{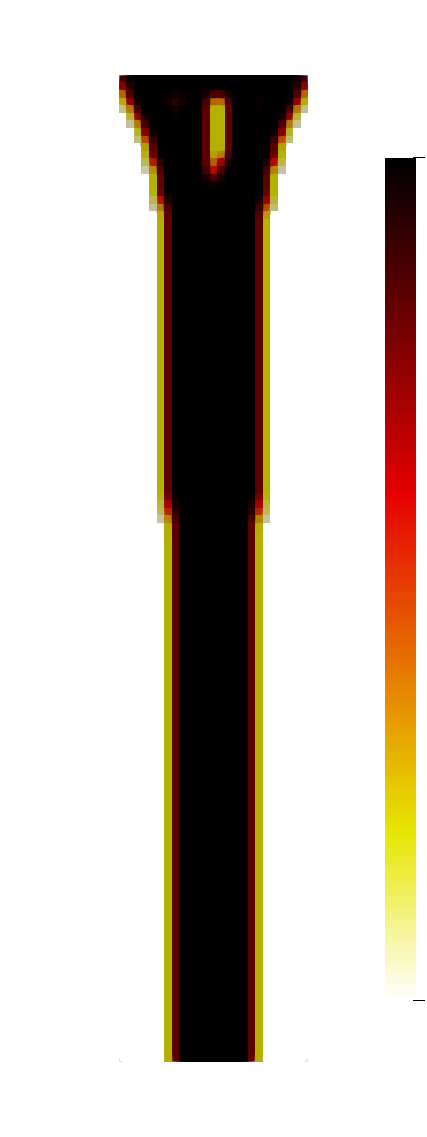}}
\begin{figure}[t]
  \centering
  \newsavebox{\simpbox}
  \savebox{\simpbox}{\includegraphics[trim={119px 75px 119px 75px},clip,height=.4\textheight]{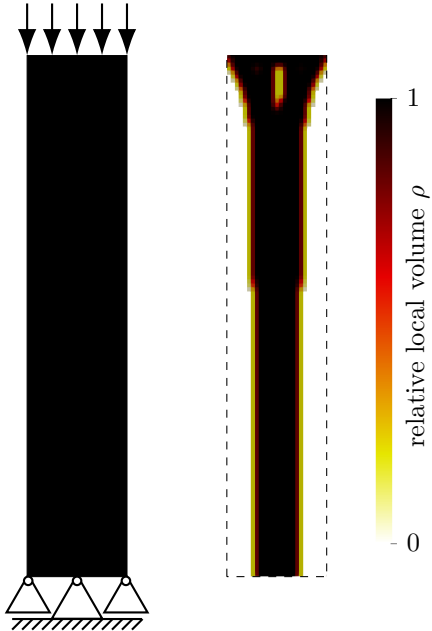}}
  \begin{tikzpicture} [
      triangle/.style={regular polygon, regular polygon sides=3},
      scale=0.8
    ]
    % design domain
    \draw[fill=black] (0,0) rectangle (\wd\simpbox,\ht\simpbox);
    % force arrows
    \foreach \xValue in {0.00,0.25,...,1.00} {
      \draw[thick,-{Latex[length=10pt, width=7pt]}] (\xValue\wd\simpbox,1.1\ht\simpbox)--(\xValue\wd\simpbox,\ht\simpbox);
    };
    % bearings
    \node[triangle,draw,thick] at (0.5pt,-11pt) {};
    \draw[fill=white,thick] (0.5pt,-2pt) circle(2pt);
    \node[triangle,draw,thick,minimum size=22pt] at (0.5\wd\simpbox,-13pt) {};
    \draw[fill=white,thick] (0.5\wd\simpbox,-2pt) circle(2pt);
    \node[triangle,draw,thick] at (\wd\simpbox-0.5pt,-11pt) {};
    \draw[fill=white,thick] (\wd\simpbox-0.5pt,-2pt) circle(2pt);
    \draw[thick] (-0.15\wd\simpbox,-20pt) -- (1.15\wd\simpbox,-20pt);
    \foreach \xValue in {-0.15,-0.05,...,1.10} {
      \draw[thick] (\xValue\wd\simpbox+5pt,-20pt)--(\xValue\wd\simpbox,-25pt);
    };
    % design
    \node[above right,inner sep=0pt,scale=0.8] at (2\wd\simpbox,0) {\usebox{\simpbox}};
    \draw[dashed] (2\wd\simpbox,0) rectangle +(\wd\simpbox,\ht\simpbox);
    % colorbar
    \node[above right,inner sep=0pt,scale=0.8] at (3.3\wd\simpbox,0) {\usebox{\simpcbbox}};
    % colorbar notation
    \node[right] at (3.3\wd\simpbox+\wd\simpcbbox,.065\ht\simpbox) {0};
    \node[right] at (3.3\wd\simpbox+\wd\simpcbbox,.92\ht\simpbox) {1};
    \node[rotate=90] at (3.3\wd\simpbox+1.5\wd\simpcbbox,.5\ht\simpbox) {relative local volume $\rho$};
  \end{tikzpicture}
  \caption{Left: macroscopic design setting with loadings and bearings. Right: the solution of a standard compliance problem \eqref{eq:complianceprob} in the design domain (dashed).}
  \label{fig:setting}
\end{figure}

Now, consider the setting shown in \cref{fig:setting}, left.
A rectangular design domain is subject to a pressure load at the top.
At the bottom rolling boundary conditions are applied, i.e., the degree of freedom in horizontal direction is fixed for all nodes at the bottom edge.
To prevent rigid body movement,  the degree of freedom in vertical direction is also fixed for the central node at the bottom edge.
This corresponds to Euler's case of a slender column with fixed/free boundary conditions, for which the buckling load is given by
\begin{equation}\label{eq:euler}
f_\text{crit}=\tfrac{\pi^2}{4 L^2}E I,
\end{equation}
where $L$ is the length of the column, $E$ is the elastic modulus and $I$ is the planar second moment of area.
The design domain has a ratio of $1:5.2$ and we discretize it with $25\times130$ bi-linear quadrilateral elements ($\mathcal{Q}_4$), which is a sufficient resolution for this type of element (compare also \cite[Fig. 1]{ferrari2019revisiting}).
\cref{fig:setting} shows the optimized design of a mechanical compliance minimization with a SIMP (Solid Isotropic Material with Penalization) material model and a lower physical design bound of $1e{-}9$.

\subsection{Pre-study: Endoskeleton versus exoskeleton}\label{ssec:parametericstudy}
\begin{figure}[t]
  \centering
  \includegraphics[trim=0 68 0 78, clip, height=.2\textheight]{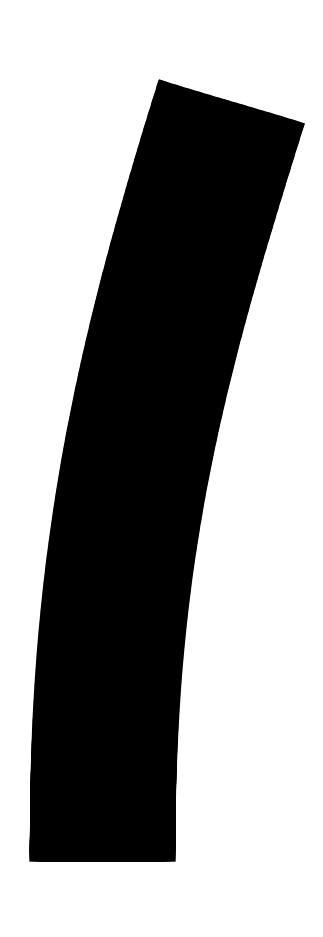}\qquad\qquad
  \includegraphics[height=.2\textheight]{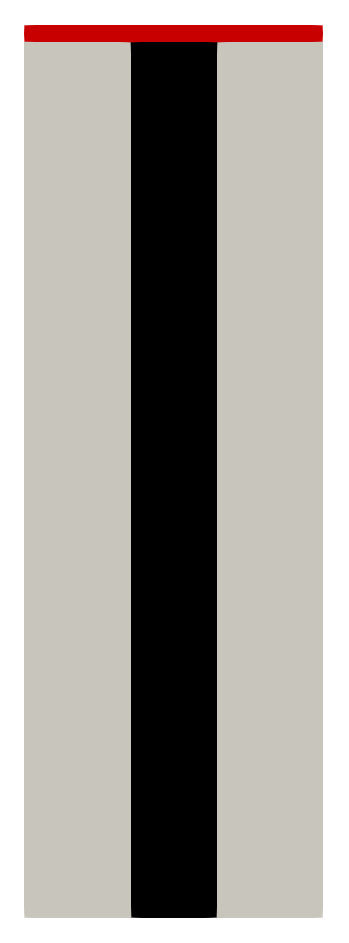}\quad
  \includegraphics[height=.2\textheight]{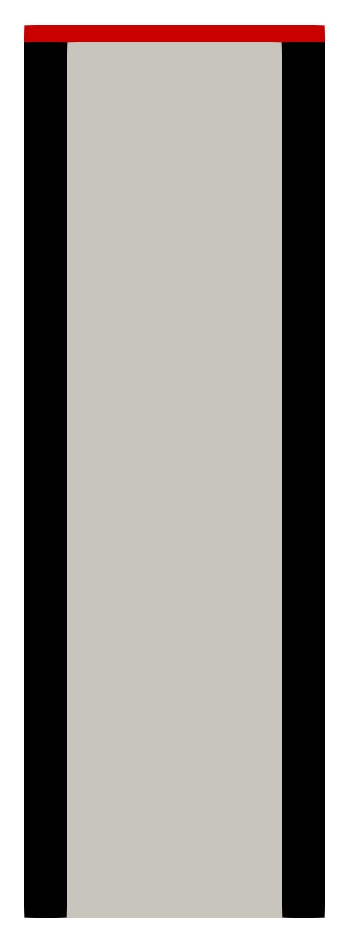}
  \caption{Left: buckling mode of a solid column. Center: column with weak (gray) material and a solid (black) reinforcement in endoskeleton configuration. Right: exoskeleton configuration. The force is applied via a stiff plate at the top (red).}
  \label{fig:endoexoskeleton}
\end{figure}

Before we look into results of the optimization problems, we perform a study to develop a better understanding of later optimized designs.
As a first step, consider $\rho=1$ on $\Omega$, i.e., solid material everywhere (\cref{fig:endoexoskeleton}, left).
Assuming plane stress conditions, a width and virtual depth of $1$\,m each, length $5.2$\,m and $E=10$\,Pa, we get $f_\text{crit}=0.076$\,N from \cref{eq:euler}.
This fits well to our numerical result of $f_\text{crit}=0.074$\,N.
Next, we compare this solid column with two structures that have the same weight each as the solid column:
weak material ($\rho=0.4$) with either a solid reinforcement ($\rho=1$) of width $0.5$\,m centered in the middle (endoskeleton, \cref{fig:endoexoskeleton}, mid) or a solid reinforcement that flanks the weak material at both sides with width $0.25$\,m (exoskeleton, \cref{fig:endoexoskeleton}, right).
The force is applied at the top boundary via a solid plate with an elastic modulus that is 100 times higher than the solid material.
As the weight is kept constant, these two structures are wider than the first one (their width is $1.75$\,m).
The critical loads are $0.065$\,N and $0.22$\,N, respectively.
Due to a larger second moment of inertia, the exoskeleton shows superior macroscopic buckling stiffness compared to the endoskeleton.
Thus, we expect designs that are optimized with respect to macroscopic buckling to exhibit an exoskeleton-like structure.

\subsection{Optimization}\label{ssec:optimization}
Next, we investigate the three different multi-objective optimization problems as given in \cref{sec:sizingopt}.
Let us briefly note that we did not run into the problem of switching eigenvalues in any of the following examples.
The material models are given by the interpolation functions from \cref{ssec:worstcasemodel} and a density filter \citep{borrvall2001topology} is applied with a radius of $1.6$ times the edge length of a finite element for regularization purpose.
We choose $\rho_\text{min}=10\%$ to obtain easily realizable lattice structures
and apply a global volume constraint with $V=50\%$ of the design domain's area.

\subsubsection*{A) Pure macroscopic load factor optimization}
Ignoring buckling on the microscopic scale, we maximize the macroscopic load factor \cref{eq:probmacrobucklingconstraint}.
The load factors for different values of the compliance constraint can be seen in \cref{fig:macroOpt}.
As reference, we include the value of a topology optimization with power law ($\rho^3$) and $\rho_\text{min}=0.001$.
Selected designs are shown in \cref{fig:macroOptDesigns}.
Design A$_1$ results from a compliance minimization without considering buckling.
Relaxing the compliance constraint while maximizing the buckling load factor, we only see a rather small improvement of up to $15\%$ in the macroscopic load factor.
This is due to design A$_1$ having already a rather good macroscopic buckling resistance, as it forms an exoskeleton.
Only in the upper part of the design domain, where the loaded edge has to be supported, is this alleviated in favor of a typical branching structure.
With a larger compliance, bound diagonal bars appear (A$_2$-A$_4$), which are known to increase macroscopic buckling stability \citep{bendsoe2003topology}.
The lower design bound of $\rho_\text{min}=10\%$ is active for all designs, see \cref{fig:macroOpt}.
We conclude from this, that the load factor could potentially be larger, i.e., better structures could be obtained, if we chose a lower $\rho_\text{min}$.
On the other hand, for very low $\rho_\text{min}$, the problem would no longer be a sizing but rather a topology optimization problem, which would require special handling (see, e.g., \cite{christensen2022multiscale}).

\pgfplotstableread[header=false]{opt_SIMP.txt}{\optsimp}
\pgfplotstablegetelem{0}{[index]3}\of{\optsimp} \pgfmathsetmacro\optsimpc{\pgfplotsretval}

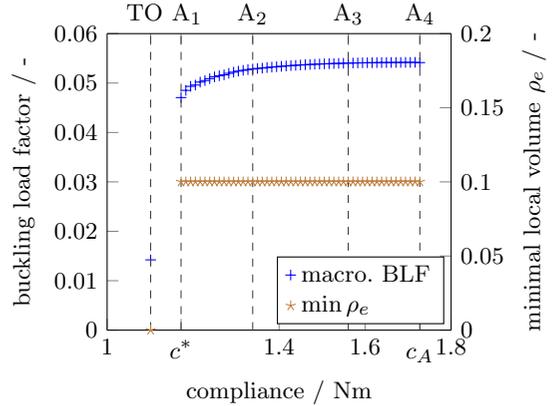
\begin{figure}[t]
  \centering
  \pgfplotstableread[header=false]{opt_macro.txt}{\optmacro}
  \pgfplotstablegetelem{0}{[index]6}\of{\optmacro} \pgfmathsetmacro\optmacroca{\pgfplotsretval}
  \pgfplotstablegetelem{15}{[index]6}\of{\optmacro} \pgfmathsetmacro\optmacrocaa{\pgfplotsretval}
  \pgfplotstablegetelem{35}{[index]6}\of{\optmacro} \pgfmathsetmacro\optmacrocaaa{\pgfplotsretval}
  \pgfplotstablegetelem{50}{[index]6}\of{\optmacro} \pgfmathsetmacro\optmacrocaaaa{\pgfplotsretval}
  \begin{tikzpicture}
    \pgfplotsset{
      width=0.6\columnwidth,
      set layers,
      scale only axis,
      xmin=1.0,
      legend style={at={(0.98,0.02)},anchor=south east,only marks}}
    \begin{axis}[
      axis x line*=bottom,
      xtick={1.0,1.4,1.6,1.8},
      extra x ticks={\optmacroca,\optmacrocaaaa},
      extra x tick labels={\phantom{$_*$}$c^*$\phantom{$_*$},\phantom{$^*$}$c_A$\phantom{$^*$}},
      xlabel=compliance / Nm,
      axis y line*=left,
      ymin=0,ymax=0.06,
      ytick distance=0.01,
      yticklabel style={/pgf/number format/.cd,fixed,precision=2},
      ylabel=buckling load factor / -,
      clip=false]
      \addplot[blue, mark=+] table [x index=6, y index=3, header=false, only marks] {\optmacro};
      \label{plot:lf_macro}
      \addplot[blue, mark=+] table [x index=3, y index=5, header=false, only marks] {\optsimp};
      \label{plot:simp_macro}
    \end{axis}
    \begin{axis}[
      axis x line*=top,
      xtick=\empty,
      extra x ticks={\optsimpc,\optmacroca,\optmacrocaa,\optmacrocaaa,\optmacrocaaaa},
      extra x tick labels={TO\phantom{$_0$}, \phantom{0}A$_1$, A$_2$, A$_3$, A$_4$},
      axis y line*=right,
      ymin=0, ymax=0.2,
      yticklabel style={/pgf/number format/.cd,fixed,precision=2},
      ylabel=minimal local volume $\rho_e$ / -]
      \addlegendimage{/pgfplots/refstyle=plot:lf_macro}\addlegendentry{macro. BLF}
      \addplot[brown, mark=star] table [x index=6, y index=0, header=false, only marks] {\optmacro};
      \addlegendentry{$\min \rho_e$}
      \addplot[brown, mark=star] table [x index=3, y index=0, header=false, only marks] {\optsimp};
      \draw[darkgray, dashed] (current axis.south-|\optsimpc, 0) -- (current axis.north-|\optsimpc,0);
      \draw[darkgray, dashed] (current axis.south-|\optmacroca, 0) -- (current axis.north-|\optmacroca,0);
      \draw[darkgray, dashed] (current axis.south-|\optmacrocaa, 0) -- (current axis.north-|\optmacrocaa,0);
      \draw[darkgray, dashed] (current axis.south-|\optmacrocaaa, 0) -- (current axis.north-|\optmacrocaaa,0);
      \draw[darkgray, dashed] (current axis.south-|\optmacrocaaaa, 0) -- (current axis.north-|\optmacrocaaaa,0);
    \end{axis}
  \end{tikzpicture}
  \caption{Optimized function values considering only macroscopic buckling subject to a compliance and a global volume constraint. As reference, we include a topology optimization result (TO) for pure compliance minimization with $\rho_\text{min} = 0.001$ with compliance $1.10$\,Nm and load factor $0.014$. Selected designs are shown in \cref{fig:macroOptDesigns}.}
  \label{fig:macroOpt}
\end{figure}
\begin{figure}[t]
  \centering
  \begin{minipage}[t]{.212\columnwidth}
    \centering
    \includegraphics[width=\textwidth]{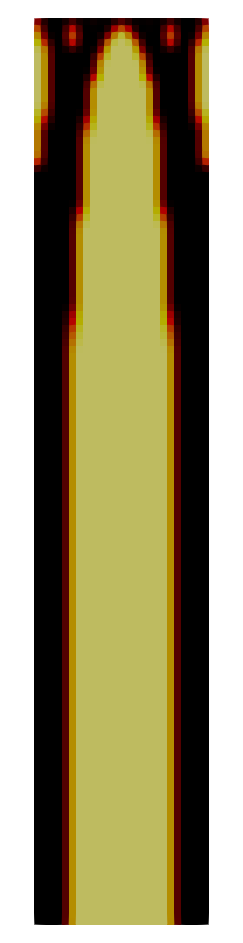}\\
    A$_1$
  \end{minipage}
  \begin{minipage}[t]{.212\columnwidth}
    \centering
    \includegraphics[width=\textwidth]{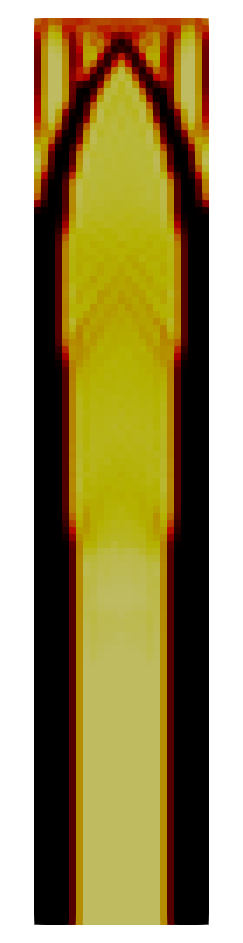}\\
    A$_2$
  \end{minipage}
  \begin{minipage}[t]{.212\columnwidth}
    \centering
    \includegraphics[width=\textwidth]{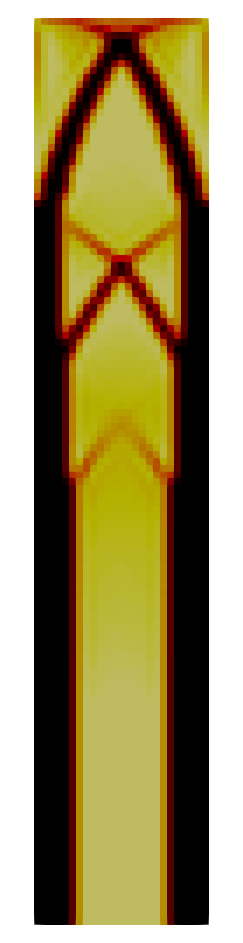}\\
    A$_3$
  \end{minipage}
  \begin{minipage}[t]{.212\columnwidth}
    \centering
    \includegraphics[width=\textwidth]{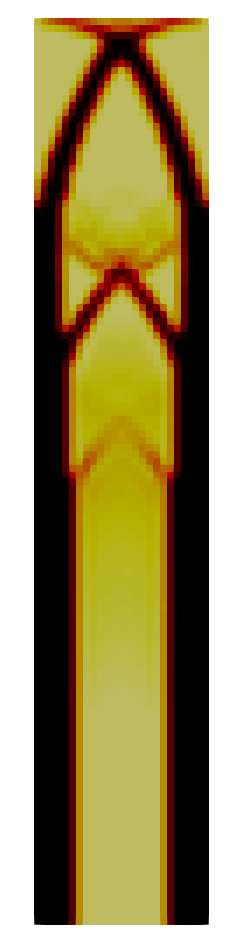}\\
    A$_4$
  \end{minipage}
  \begin{minipage}[t]{.1\columnwidth}
    \centering
    \begin{tikzpicture} [scale=0.7]
      % colorbar
      \node[above right,inner sep=0pt,scale=0.7] at (0,0) {\usebox{\simpcbbox}};
      % colorbar notation
      \node[right] at (0+\wd\simpcbbox,.065\ht\simpcbbox) {0};
      \node[right] at (0+\wd\simpcbbox,.92\ht\simpcbbox) {1};
      \node[rotate=90] at (0+1.5\wd\simpcbbox,.5\ht\simpcbbox) {local volume / -};
    \end{tikzpicture}
  \end{minipage}
  \caption{Optimized designs for the marked data in \cref{fig:macroOpt}. From left to right the compliance constraint is relaxed and macroscopic buckling stiffness increases through appearance of reinforcing, diagonal structures.}
  \label{fig:macroOptDesigns}
\end{figure}
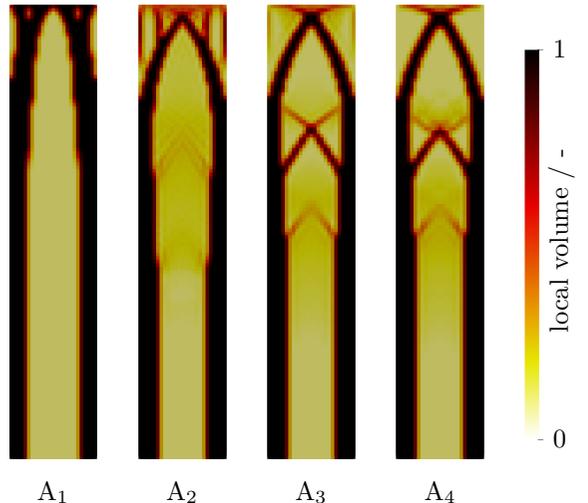

\subsubsection*{B) Pure microscopic load factor optimization}
In the second optimization problem, we ignore macroscopic buckling and instead maximize the smallest microscopic load factor of all finite elements \cref{eq:probmicrobucklingconstraint} subject to a compliance constraint.
The microscopic buckling load factors are approximated by our worst-case model \cref{eq:intbuckling}.
In \cref{fig:microOpt}, a substantial improvement of the load factor can be seen when the compliance constraint is relaxed. 
The optimal solution for the problem without compliance constraint (design B$_5$) is a fully homogeneous design (see \cref{fig:microOptDesigns}).
Design B$_5$ exhibits homogeneous stress under the given boundary conditions, which results in a homogeneous microscopic load factor.
Thus, the further we relax the compliance constraint, the more homogeneous the optimized design becomes.
First, the branching structure at the top of design B$_1$=A$_1$ is replaced by lattice with high density (B$_2$,B$_3$), then the load carrying skeleton vanishes (B$_4$,B$_5$).
The jumps in the minimal volume in \cref{fig:microOpt} can be explained by design changes and the discretization of the design domain, e.g., between B$_3$ and B$_4$, the solid parts on the left and right edge each become one finite element thinner, which leads to a different stress distribution in the whole structure.

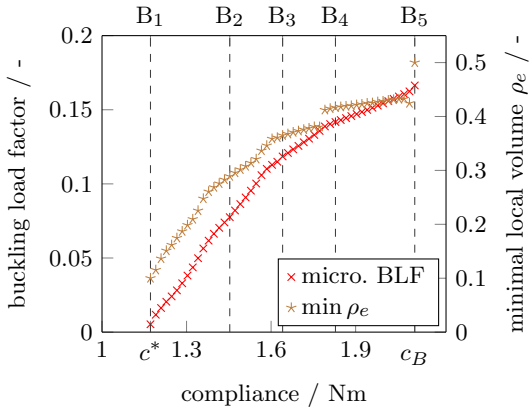
\begin{figure}[t]
  \centering
  \pgfplotstableread[header=false]{opt_micro.txt}{\optmicro}
  \pgfplotstablegetelem{0}{[index]3}\of{\optmicro} \pgfmathsetmacro\optmicrocb{\pgfplotsretval}
  \pgfplotstablegetelem{15}{[index]3}\of{\optmicro} \pgfmathsetmacro\optmicrocbb{\pgfplotsretval}
  \pgfplotstablegetelem{25}{[index]3}\of{\optmicro} \pgfmathsetmacro\optmicrocbbb{\pgfplotsretval}
  \pgfplotstablegetelem{35}{[index]3}\of{\optmicro} \pgfmathsetmacro\optmicrocbbbb{\pgfplotsretval}
  \pgfplotstablegetelem{50}{[index]3}\of{\optmicro} \pgfmathsetmacro\optmicrocbbbbb{\pgfplotsretval}
  \begin{tikzpicture}
    \pgfplotsset{
      width=0.6\columnwidth,
      set layers,
      scale only axis,
      xmin=1.0,
      legend style={at={(0.98,0.02)},anchor=south east,only marks}}
    \begin{axis}[
      axis x line*=bottom,
      xtick={1.0,1.3,1.6,1.9},
      extra x ticks={\optmicrocb,\optmicrocbbbbb},
      extra x tick labels={\phantom{$_*$}$c^*$\phantom{$_*$},\phantom{$^*$}$c_B$\phantom{$^*$}},
      xlabel=compliance / Nm,
      axis y line*=left,
      ymin=0,ymax=0.2,
      ytick distance=0.05,
      yticklabel style={/pgf/number format/.cd,fixed,precision=2},
      ylabel=buckling load factor / -,
      clip=false]
      \addplot[red, mark=x] table [x index=3, y index=1, header=false, only marks] {\optmicro};
      \label{plot:lf_micro}
    \end{axis}
    \begin{axis}[
      axis x line*=top,
      xtick=\empty,
      extra x ticks={\optmicrocb,\optmicrocbb,\optmicrocbbb,\optmicrocbbbb,\optmicrocbbbbb},
      extra x tick labels={B$_1$, B$_2$, B$_3$, B$_4$, B$_5$},
      axis y line*=right,
      ymin=0, ymax=0.55,
      ytick distance=0.1,
      yticklabel style={/pgf/number format/.cd,fixed,precision=2},
      ylabel=minimal local volume $\rho_e$ / -]
      \addlegendimage{/pgfplots/refstyle=plot:lf_micro}\addlegendentry{micro. BLF}
      \addplot[brown, mark=star] table [x index=3, y index=0, header=false, only marks] {\optmicro};
      \addlegendentry{$\min \rho_e$}
      \draw[darkgray, dashed] (current axis.south-|\optmicrocb, 0) -- (current axis.north-|\optmicrocb,0);
      \draw[darkgray, dashed] (current axis.south-|\optmicrocbb, 0) -- (current axis.north-|\optmicrocbb,0);
      \draw[darkgray, dashed] (current axis.south-|\optmicrocbbb, 0) -- (current axis.north-|\optmicrocbbb,0);
      \draw[darkgray, dashed] (current axis.south-|\optmicrocbbbb, 0) -- (current axis.north-|\optmicrocbbbb,0);
      \draw[darkgray, dashed] (current axis.south-|\optmicrocbbbbb, 0) -- (current axis.north-|\optmicrocbbbbb,0);
    \end{axis}
  \end{tikzpicture}
  \caption{Optimized function values considering only microscopic buckling subject to a compliance and a global volume constraint. The optimized designs show increasing minimal local volume (see also \cref{fig:microOptDesigns}), which yields better local buckling resistance.}
  \label{fig:microOpt}
\end{figure}
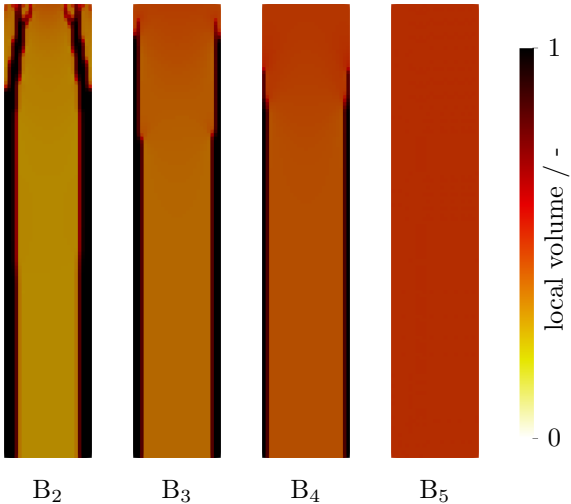
\begin{figure}[t]
  \centering
  \begin{minipage}[t]{.212\columnwidth}
    \centering
    \includegraphics[width=\textwidth]{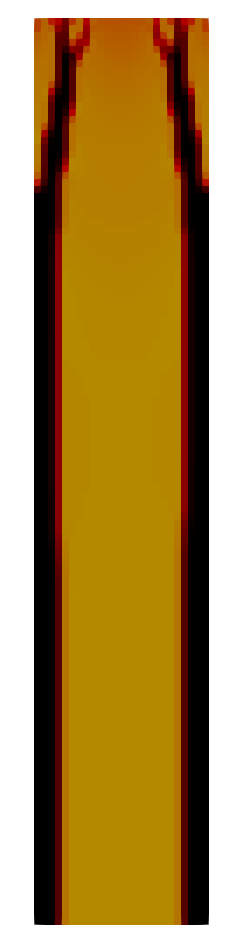}\\
    B$_2$
  \end{minipage}
  \begin{minipage}[t]{.212\columnwidth}
    \centering
    \includegraphics[width=\textwidth]{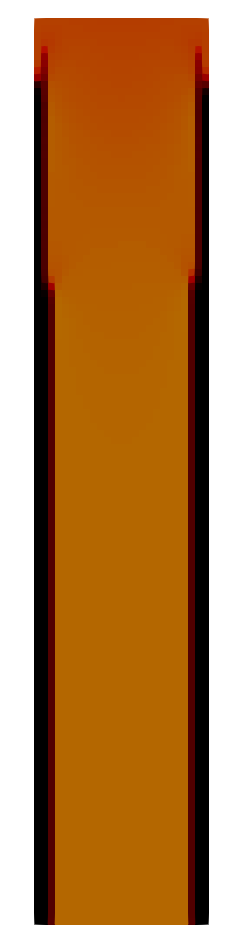}\\
    B$_3$
  \end{minipage}
  \begin{minipage}[t]{.212\columnwidth}
    \centering
    \includegraphics[width=\textwidth]{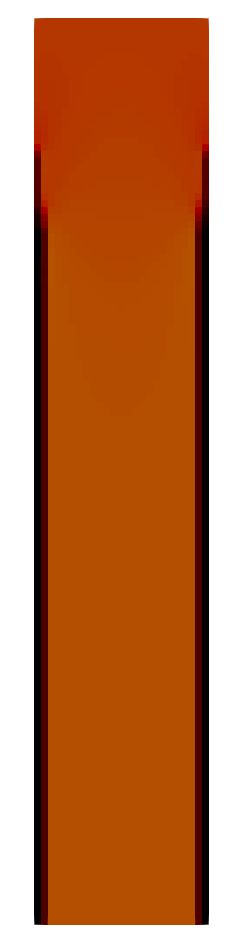}\\
    B$_4$
  \end{minipage}
  \begin{minipage}[t]{.212\columnwidth}
    \centering
    \includegraphics[width=\textwidth]{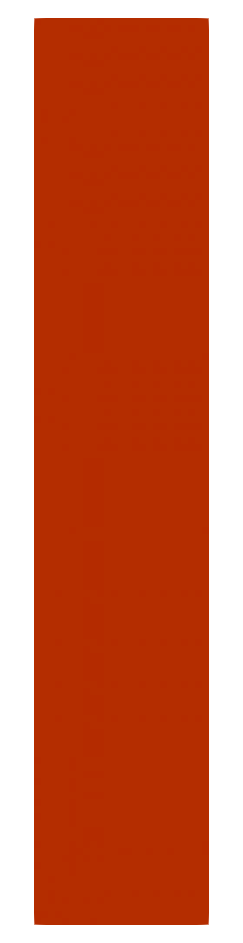}\\
    B$_5$
  \end{minipage}
  \begin{minipage}[t]{.1\columnwidth}
    \centering
    \begin{tikzpicture} [scale=0.7]
      % colorbar
      \node[above right,inner sep=0pt,scale=0.7] at (0,0) {\usebox{\simpcbbox}};
      % colorbar notation
      \node[right] at (0+\wd\simpcbbox,.065\ht\simpcbbox) {0};
      \node[right] at (0+\wd\simpcbbox,.92\ht\simpcbbox) {1};
      \node[rotate=90] at (0+1.5\wd\simpcbbox,.5\ht\simpcbbox) {local volume / -};
    \end{tikzpicture}
  \end{minipage}
  \caption{Optimized designs for the marked data in \cref{fig:microOpt}. Design B$_1$ equals A$_1$ in \cref{fig:macroOptDesigns}. B$_5$ is pure maximization of microscopic buckling without compliance constraint and yields a homogeneous design.}
  \label{fig:microOptDesigns}
\end{figure}

\subsubsection*{C) Simultaneous optimization of pure macro- and microscopic load factors}
The obtained values for the simultaneous optimization of both  macroscopic and microscopic buckling load factors \cref{prob:macroopt}-\cref{eq:probcomplianceconstraint} can be seen in \cref{fig:macroMicroOpt}.
As reference, the values of a topology optimization (TO) with $\rho=0.001$ are given.
To achieve a stiff (small compliance) design, thick solid structures are needed.
Little material is left for the lattice part, which results in low local volume and a small minimal microscopic buckling load factor.
Hence, only \cref{eq:probmicrobucklingconstraint,eq:probcomplianceconstraint} are active and \cref{eq:probmacrobucklingconstraint} remains inactive.
For less restrictive (larger) compliance bounds, less solid material is necessary and material is redistributed to the lattice region (C$_2$).
Thus, the microscopic buckling load factor can be improved.
When it reaches the value of the macroscopic one, \cref{eq:probmacrobucklingconstraint} becomes active (C$_3$).
Therefore, raising only the microscopic buckling load factor further, i.e., steering towards homogeneous design as in B, yields no improvement in the objective, as the macroscopic load factor will define the value of the slack variable.
For this reason, both micro- and macroscopic buckling load factors are raised simultaneously:
the first by increasing the lattice density especially in the upper region of the design domain and the second by stiffening the exoskeleton (C$_4$,C$_5$).

The decreasing minimal local volume in \cref{fig:macroMicroOpt} for more relaxed compliance appears non-intuitive.
In \cref{fig:macroMicroOptDesigns}, staircase structures on the solid parts can be seen due to discretization, see, e.g., C$_5$.
The resulting local stress field allows a slightly lower volume for individual finite elements, while preserving the microscopic buckling load factor.

\pgfplotstableread[header=false]{opt_macromicro.txt}{\optmaicro}
\pgfplotstablegetelem{35}{[index]7}\of{\optmaicro} \pgfmathsetmacro\optmaicroccccc{\pgfplotsretval}

\begin{figure}[t]
  \centering
  \pgfplotstablegetelem{0}{[index]7}\of{\optmaicro} \pgfmathsetmacro\optmaicrocc{\pgfplotsretval}
  \pgfplotstablegetelem{15}{[index]7}\of{\optmaicro} \pgfmathsetmacro\optmaicroccc{\pgfplotsretval}
  \pgfplotstablegetelem{25}{[index]7}\of{\optmaicro} \pgfmathsetmacro\optmaicrocccc{\pgfplotsretval}
  \pgfplotstablegetelem{50}{[index]7}\of{\optmaicro} \pgfmathsetmacro\optmaicrocccccc{\pgfplotsretval}
  \begin{tikzpicture}
    \pgfplotsset{
      width=0.6\columnwidth,
      set layers,
      scale only axis,
      xmin=1.0,
      legend style={at={(0.98,0.02)},anchor=south east,only marks}}
    \begin{axis}[
      axis x line*=bottom,
      xtick={1.0,1.3,1.4},
      extra x ticks={\optmaicrocc,\optmaicrocccccc},
      extra x tick labels={\phantom{$_*$}$c^*$\phantom{$_*$},\phantom{$^*$}$c_C$\phantom{$^*$}},
      xlabel=compliance / Nm,
      axis y line*=left,
      ymin=0,ymax=0.06,
      ytick distance=0.01,
      yticklabel style={/pgf/number format/.cd,fixed,precision=2},
      ylabel=buckling load factor / -,
      clip=false]
      \addplot[blue, mark=+] table [x index=7, y index=4, header=false, only marks] {\optmaicro};
      \label{plot:lf_maicroma}
      \addplot[red, mark=x] table [x index=7, y index=1, header=false, only marks] {\optmaicro};
      \label{plot:lf_maicromi}
      \addplot[blue, mark=+] table [x index=3, y index=5, header=false, only marks] {\optsimp};
      \label{plot:simp_maicro}
    \end{axis}
    \begin{axis}[
      axis x line*=top,
      xtick=\empty,
      extra x ticks={\optsimpc, \optmaicrocc,\optmaicroccc,\optmaicrocccc,\optmaicroccccc,\optmaicrocccccc},
      extra x tick labels={TO\phantom{$_0$}, C$_1$, C$_2$, C$_3$, C$_4$, C$_5$},
      axis y line*=right,
      ymin=0, ymax=0.3,
      ytick distance=0.05,
      yticklabel style={/pgf/number format/.cd,fixed,precision=2},
      ylabel=minimal local volume $\rho_e$ / -]
      \addlegendimage{/pgfplots/refstyle=plot:lf_maicroma}\addlegendentry{macro. BLF}
      \addlegendimage{/pgfplots/refstyle=plot:lf_maicromi}\addlegendentry{micro. BLF}
      \addplot[brown, mark=star] table [x index=7, y index=0, header=false, only marks] {\optmaicro};
      \addlegendentry{$\min \rho_e$}
      \addplot[brown, mark=star] table [x index=3, y index=0, header=false, only marks] {\optsimp};
      \draw[darkgray, dashed] (current axis.south-|\optsimpc, 0) -- (current axis.north-|\optsimpc,0);
      \draw[darkgray, dashed] (current axis.south-|\optmaicrocc, 0) -- (current axis.north-|\optmaicrocc,0);
      \draw[darkgray, dashed] (current axis.south-|\optmaicroccc, 0) -- (current axis.north-|\optmaicroccc,0);
      \draw[darkgray, dashed] (current axis.south-|\optmaicrocccc, 0) -- (current axis.north-|\optmaicrocccc,0);
      \draw[darkgray, dashed] (current axis.south-|\optmaicroccccc, 0) -- (current axis.north-|\optmaicroccccc,0);
      \draw[darkgray, dashed] (current axis.south-|\optmaicrocccccc, 0) -- (current axis.north-|\optmaicrocccccc,0);
    \end{axis}
  \end{tikzpicture}
  \caption{Optimized function values of concurrent load factor maximization of pure macro- and pure microscopic buckling with compliance and volume constraint \cref{prob:macroopt}-\cref{eq:probcomplianceconstraint}. Up to compliance 1.33, only the microscopic buckling load factor is active. For higher compliance values, the macroscopic buckling load factor dominates. Selected designs are shown in \cref{fig:macroMicroOptDesigns}.}
  \label{fig:macroMicroOpt}
\end{figure}
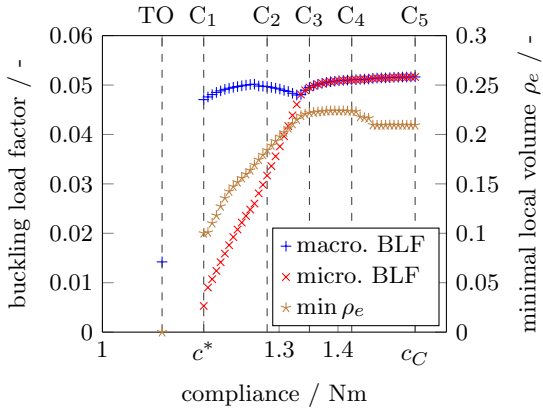
\begin{figure}[t]
  \centering
  \begin{minipage}[t]{.212\columnwidth}
    \centering
    \includegraphics[width=\textwidth]{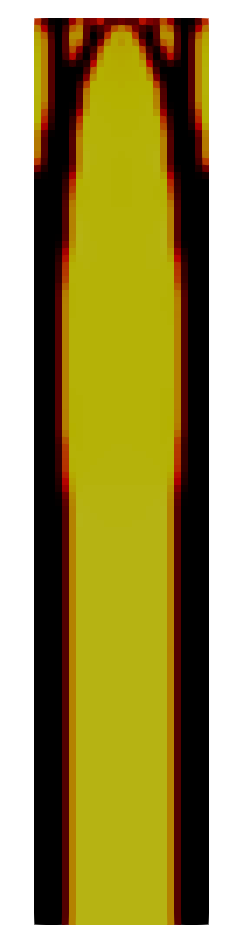}\\
    C$_2$
  \end{minipage}
  \begin{minipage}[t]{.212\columnwidth}
    \centering
    \includegraphics[width=\textwidth]{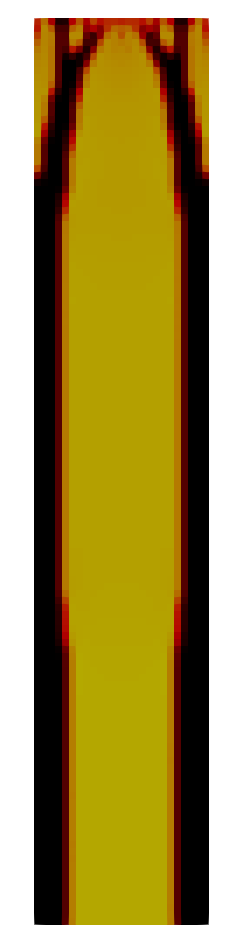}\\
    C$_3$
  \end{minipage}
  \begin{minipage}[t]{.212\columnwidth}
    \centering
    \includegraphics[width=\textwidth]{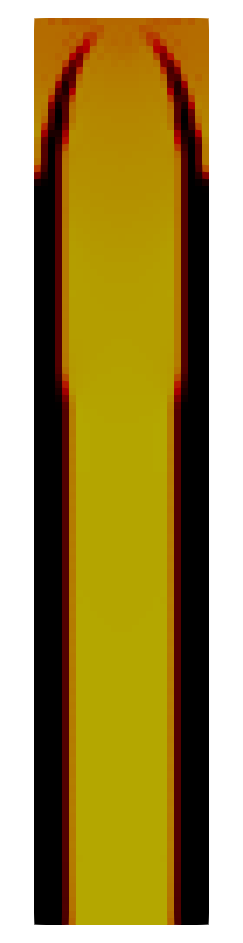}\\
    C$_4$
  \end{minipage}
  \begin{minipage}[t]{.212\columnwidth}
    \centering
    \begin{tikzpicture}[
        dot/.style = {circle, thick, minimum size=#1, inner sep=0pt, outer sep=0pt},
      ]
      \newsavebox{\mybox}
      \savebox{\mybox}{\includegraphics[width=\textwidth]{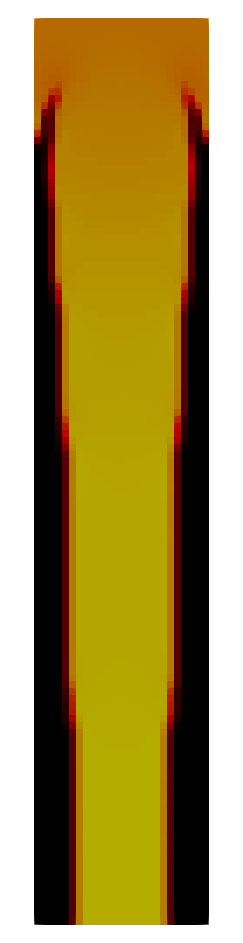}}
      \node[above right,inner sep=0pt] at (0,0) {\usebox{\mybox}};
      \draw[white,line width=1pt] (.25\wd\mybox,.68\ht\mybox) circle [x radius=.1\wd\mybox, y radius=.2\wd\mybox];
      \draw[white,line width=1pt] (.28\wd\mybox,.54\ht\mybox) circle [x radius=.1\wd\mybox, y radius=.2\wd\mybox];
      \draw[white,line width=1pt] (.31\wd\mybox,.25\ht\mybox) circle [x radius=.1\wd\mybox, y radius=.2\wd\mybox];
    \end{tikzpicture}\\
    C$_5$
  \end{minipage}
  \begin{minipage}[t]{.1\columnwidth}
    \centering
    \begin{tikzpicture} [scale=0.7]
      % colorbar
      \node[above right,inner sep=0pt,scale=0.7] at (0,0) {\usebox{\simpcbbox}};
      % colorbar notation
      \node[right] at (0+\wd\simpcbbox,.065\ht\simpcbbox) {0};
      \node[right] at (0+\wd\simpcbbox,.92\ht\simpcbbox) {1};
      \node[rotate=90] at (0+1.5\wd\simpcbbox,.5\ht\simpcbbox) {local volume / -};
    \end{tikzpicture}
  \end{minipage}
  \caption{Optimized designs for the marked data in \cref{fig:macroMicroOpt}. Design C$_1$ equals A$_1$ in \cref{fig:macroOptDesigns}. To increase microscopic buckling resistance, the optimizer reduces lattice porosity. No diagonal bars appear as they do in the pure macroscopic buckling optimization, since the lattice already provides good macroscopic buckling resistance. Due to discretization, steps can be seen on the solid parts, as marked in C$_5$ by white ellipses.}
  \label{fig:macroMicroOptDesigns}
\end{figure}
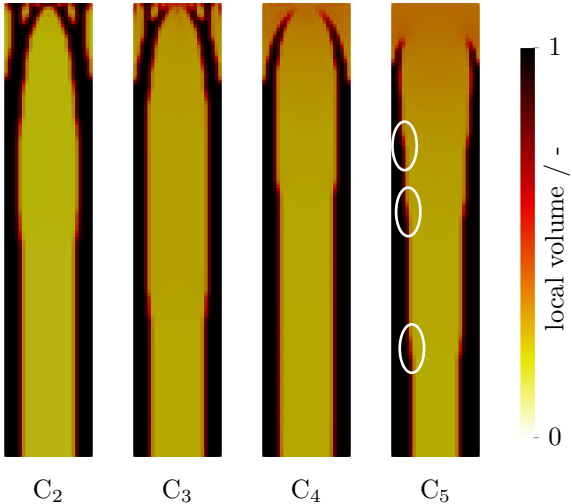

\section{Dehomogenization and validation}\label{sec:validation}%
In this section, we want to give two distinct comparisons:
First, we compare the predicted buckling behavior with high-resolution numerical analyses of dehomogenized designs.
Second, we validate the worst-case model against a precise microscopic load factor evaluation.

\subsection{Performance of dehomogenized designs}\label{ssec:dehomogenization}
The number of lattice cells for dehomogenization of optimized results is chosen independently of the finite element resolution of the macroscopic model (\cf \cref{fig:setting}).
To realize this, we proceed as follows: First, the optimized density field $\rho$ is interpolated and the number of lattice cells is chosen.
Then, an evaluation of this field at each lattice cell's center defines the width of the corresponding lattice struts.
Note that our dehomogenization procedure is volume preserving, i.e., all dehomogenized designs have a volume of $V=50\%$ of the design domain's area.
Dehomogenized designs are shown in \cref{fig:dehomdisplacement,fig:dehombuckling}.

\subsubsection*{Influence of cell size}
We compare dehomogenizations with different cell sizes for design C$_4$ from \cref{fig:macroMicroOptDesigns}.
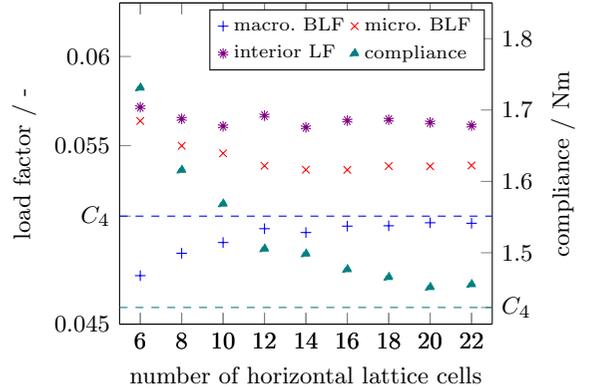
\begin{figure}[t]
  \pgfplotstableread[header=false]{cellsize.txt}{\cellsize}
  \centering
  \pgfplotstablegetelem{35}{[index]4}\of{\optmaicro} \pgfmathsetmacro\optmaicromacccc{\pgfplotsretval}
  \begin{tikzpicture}
    \pgfplotsset{
      width=0.65\columnwidth,
      xmin=5,xmax=23,
      xtick distance=2,
      set layers,
      scale only axis,
      legend columns=2,
      legend style={font=\footnotesize,at={(0.98,0.98)},anchor=north east}}
    \begin{axis}[
      xlabel=number of horizontal lattice cells,
      axis y line*=left,
      ymin=0.045,ymax=0.063,
      ytick={0.045,0.055,0.06,0.065},
      extra y ticks={\optmaicromacccc},
      extra y tick labels={${C_4}$},
      yticklabel style={/pgf/number format/.cd,fixed,precision=3},
      ylabel=load factor / -]
      \addplot[blue, mark=+, only marks] table [x index=10, y index=2, header=false] {\cellsize};
      \label{plot:macro_lf}
      \addplot[red, mark=x, only marks] table [x index=10, y index=3, header=false] {\cellsize};
      \label{plot:micro_lf1}
      \addplot[violet, mark=10-pointed star, only marks] table [x index=10, y index=4, header=false] {\cellsize};
      \label{plot:micro_lf2}
      \addplot[blue, domain=5:23, mark=none, samples=2, dashed] {\optmaicromacccc};
    \end{axis}
    \begin{axis}[
      axis y line*=right,
      ymin=1.4,ymax=1.85,
      ytick={1.5,1.6,1.7,1.8},
      extra y ticks={\optmaicroccccc},
      extra y tick labels={${C_4}$},
      yticklabel style={/pgf/number format/.cd,fixed,precision=2},
      ylabel=compliance / Nm]
      \addlegendimage{/pgfplots/refstyle=plot:macro_lf}\addlegendentry{macro. BLF}
      \addlegendimage{/pgfplots/refstyle=plot:micro_lf1}\addlegendentry{micro. BLF}
      \addlegendimage{/pgfplots/refstyle=plot:micro_lf2}\addlegendentry{interior LF}
      \addplot[teal, dashed, mark=triangle*, only marks] table [x index=10, y index=1, header=false] {\cellsize};
      \addlegendentry{compliance}
      \addplot[teal, domain=5:23, mark=none, samples=2, dashed] {\optmaicroccccc};
    \end{axis}
  \end{tikzpicture}
  \caption{Dehomogenization of design C$_4$ with increasing number of lattice cells. For the interior microscopic load factor, we searched for the first buckling mode that has no deflection at the boundary.
  The macroscopic buckling load factor matches our prediction well; the microscopic and the interior load factors are better than predicted by the worst-case model (\cf \cref{fig:macroMicroOpt}).}
  \label{fig:cellsize}
\end{figure}
We investigate different buckling load factors:
the one associated with a macroscopic buckling mode, i.e., deflection of the structure as a whole (low-frequency), and the one associated with a microscopic mode, i.e., deflection of the lattice (high-frequency).
As expected, the predicted macroscopic buckling load factor is met better with a higher number of lattice cells (see \cref{fig:cellsize}).
The microscopic buckling load factor stems from modes at the boundary (\cref{fig:dehombuckling}).
These modes cannot be approximated well by homogenization theory, which ignores boundary effects.
We therefore also search for the smallest load factor associated with an interior mode, i.e., a mode, that does not exhibit deflection at the structure's boundary (\cf \cref{fig:dehombuckling}).
In order to avoid buckling of the microstructure occurring at the structural boundary, a coating strategy as suggested by \cite{christensen2022multiscale} may be pursued.

In the considered example, the cell size has almost no influence on this interior load factor.
This might differ for other examples.
Zooms for interior microscopic buckling modes are given in \cref{fig:dehombucklingzoom}.

\begin{figure}[t]
  \centering
  \includegraphics[width=.3\columnwidth]{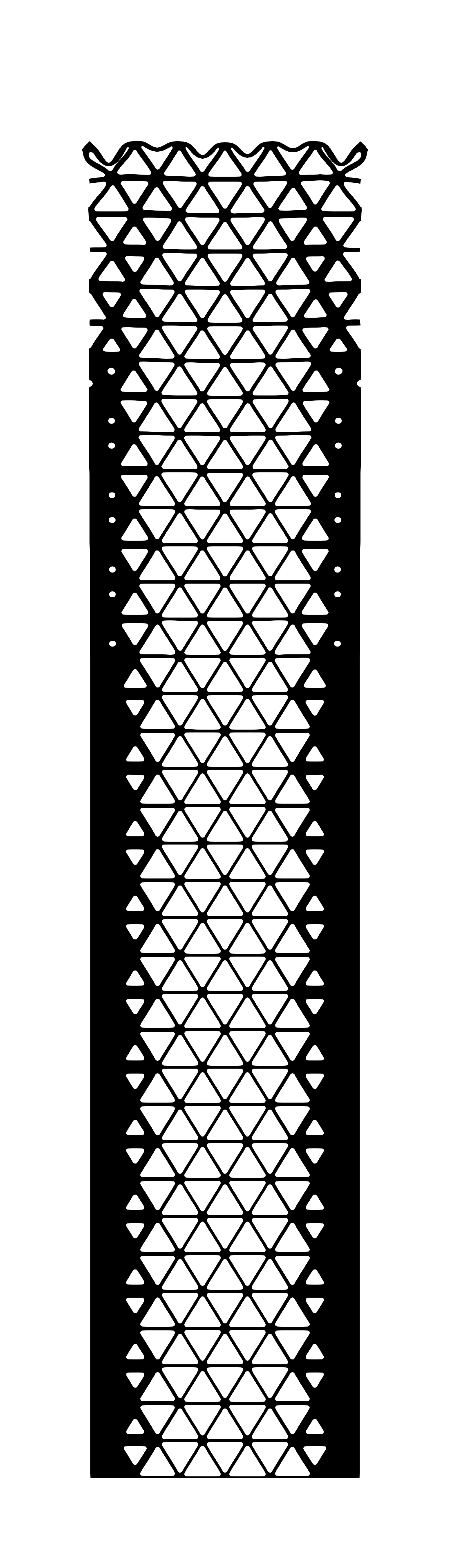}\quad
  \includegraphics[width=.3\columnwidth]{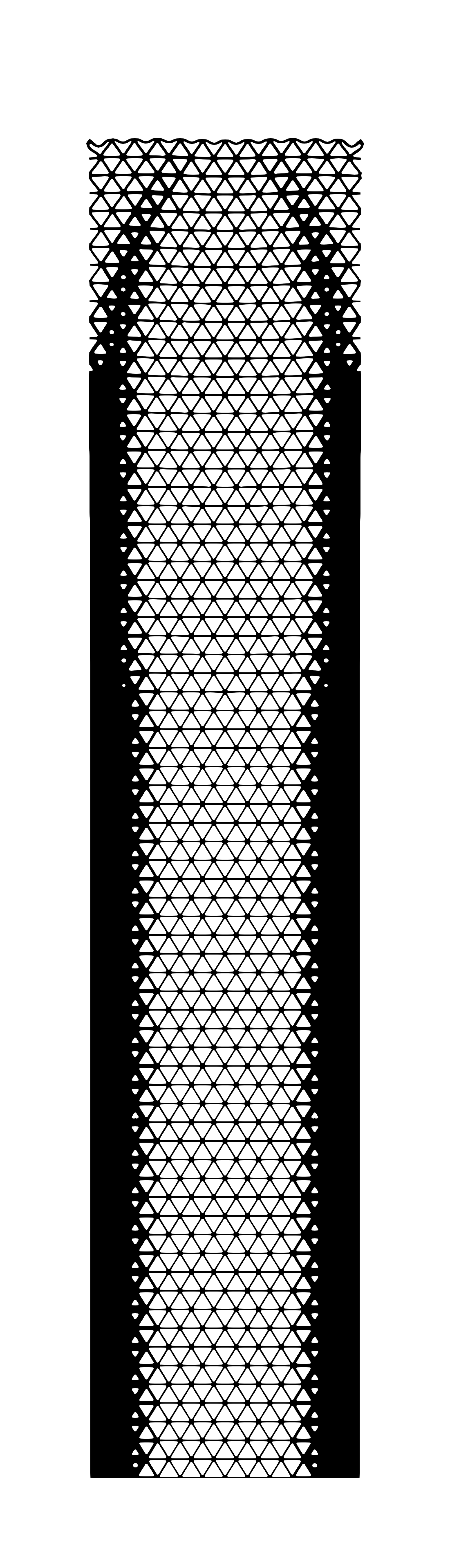}
  \caption{Dehomogenization of design C$_4$ with $6$ (left) and $12$ (right) cells in horizontal direction with visualized pre-buckling displacement. The horizontal struts at the top face where the pressure load is applied exhibit sag, which is reduced for a higher number of cells.}
  \label{fig:dehomdisplacement}
\end{figure}

As can be seen in \cref{fig:cellsize}, the compliance tends to decrease with smaller cell size.
The observed increase for 22 lattice cells compared to 20 is presumably caused by discretization effects.
\cref{fig:dehomdisplacement} shows that the displacement under the applied pressure load has a wave shape, where horizontal lattice struts at the top face sag between their diagonal supports.
The more cells we have, the better these struts are supported, and thus the compliance is reduced.

\begin{figure}
  \centering
  \pdfpxdimen=\dimexpr 1 in/72\relax
  \includegraphics[trim={450px 0 200px 0},clip,angle=90,width=.43\columnwidth]{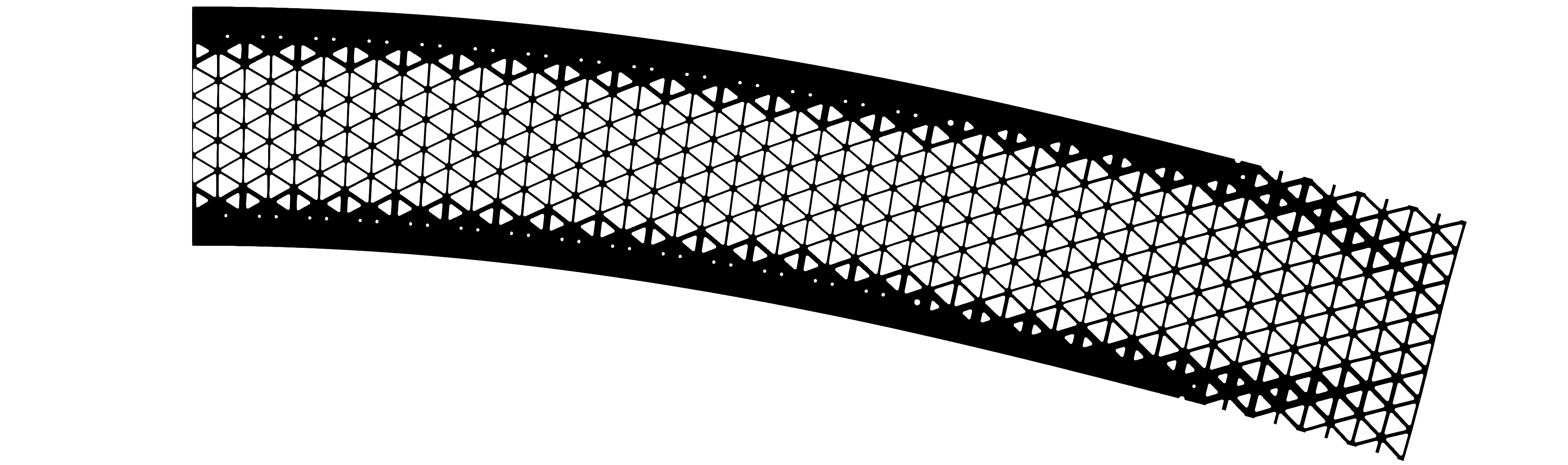}\quad
  \includegraphics[trim={450px 520px 200px 0},clip,angle=90,width=.23\columnwidth]{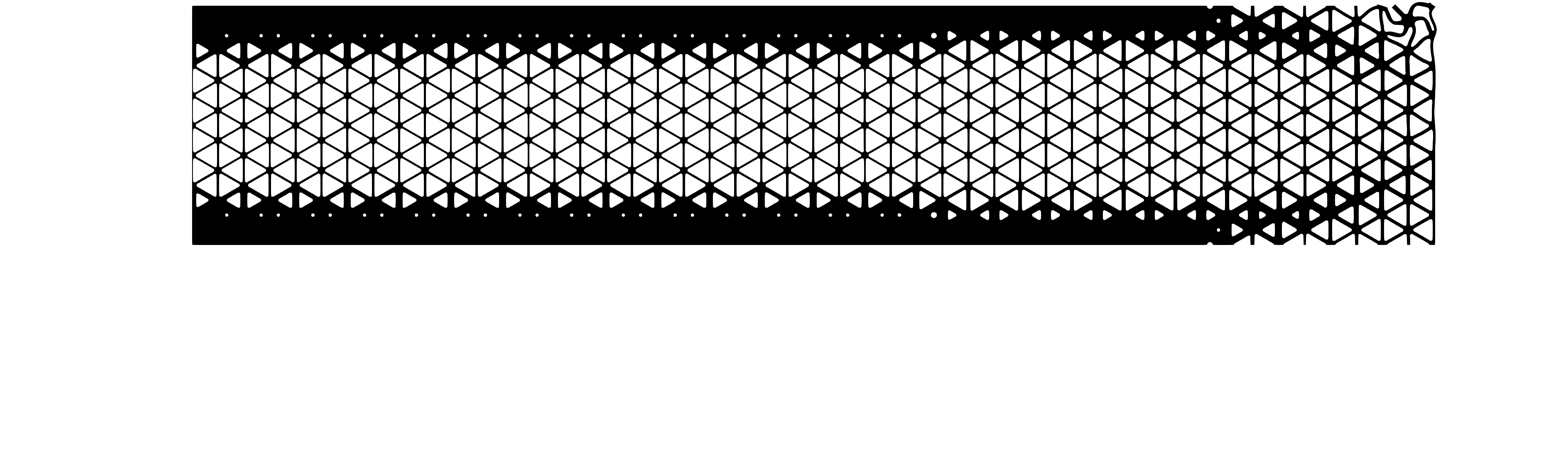}\quad
  \includegraphics[trim={450px 520px 200px 0},clip,angle=90,width=.23\columnwidth]{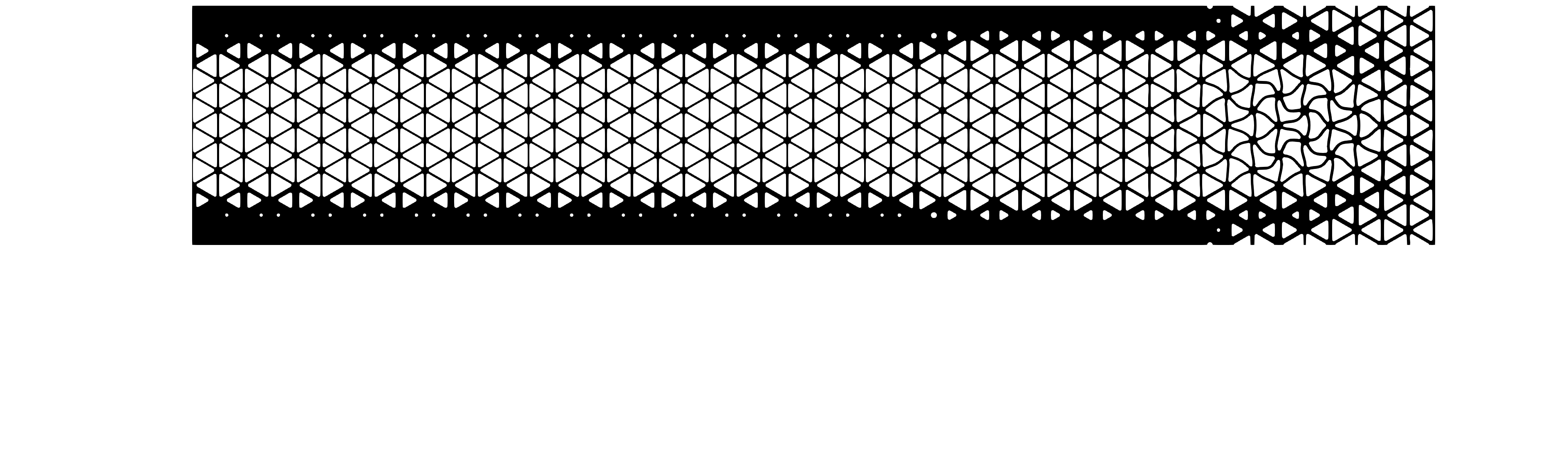}
  \caption{Different buckling modes for design C$_4$ dehomogenized with eight lattice cells (\cf \cref{fig:cellsize}). Left: first macroscopic mode. Center: first microscopic mode. Right: first interior mode.}
  \label{fig:dehombuckling}
\end{figure}
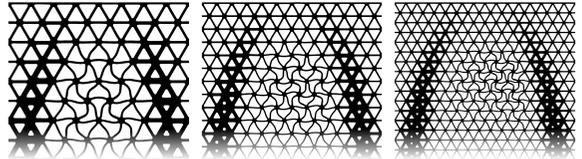
\begin{figure}[t]
  \centering
  \pdfpxdimen=\dimexpr 1 in/72\relax
  \newsavebox{\valpfem}
  \savebox{\valpfem}{\includegraphics[trim={300px 0 300px 0},clip,width=.32\columnwidth]{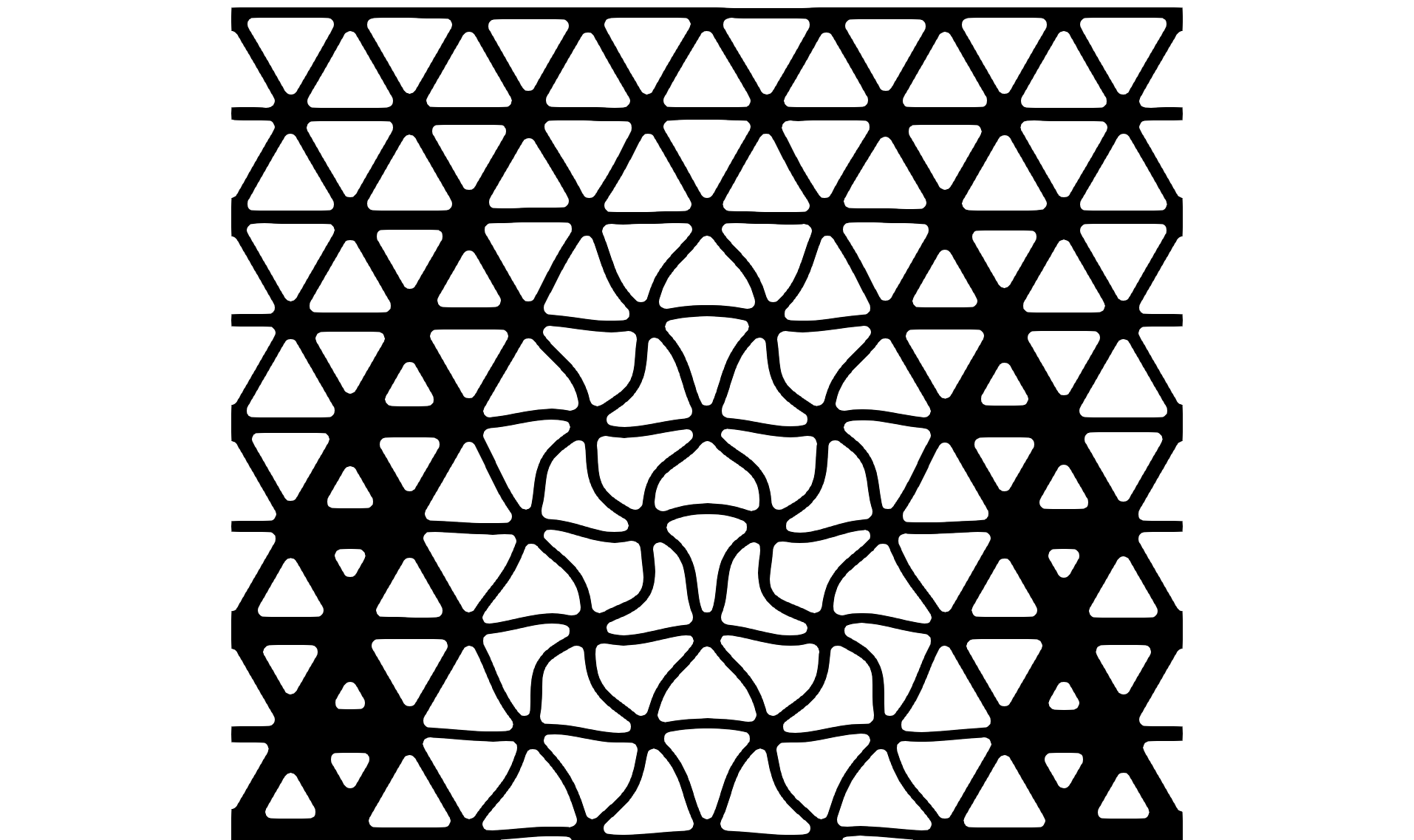}}
  \newsavebox{\valpfemm}
  \savebox{\valpfemm}{\includegraphics[trim={300px 0 300px 0},clip,width=.32\columnwidth]{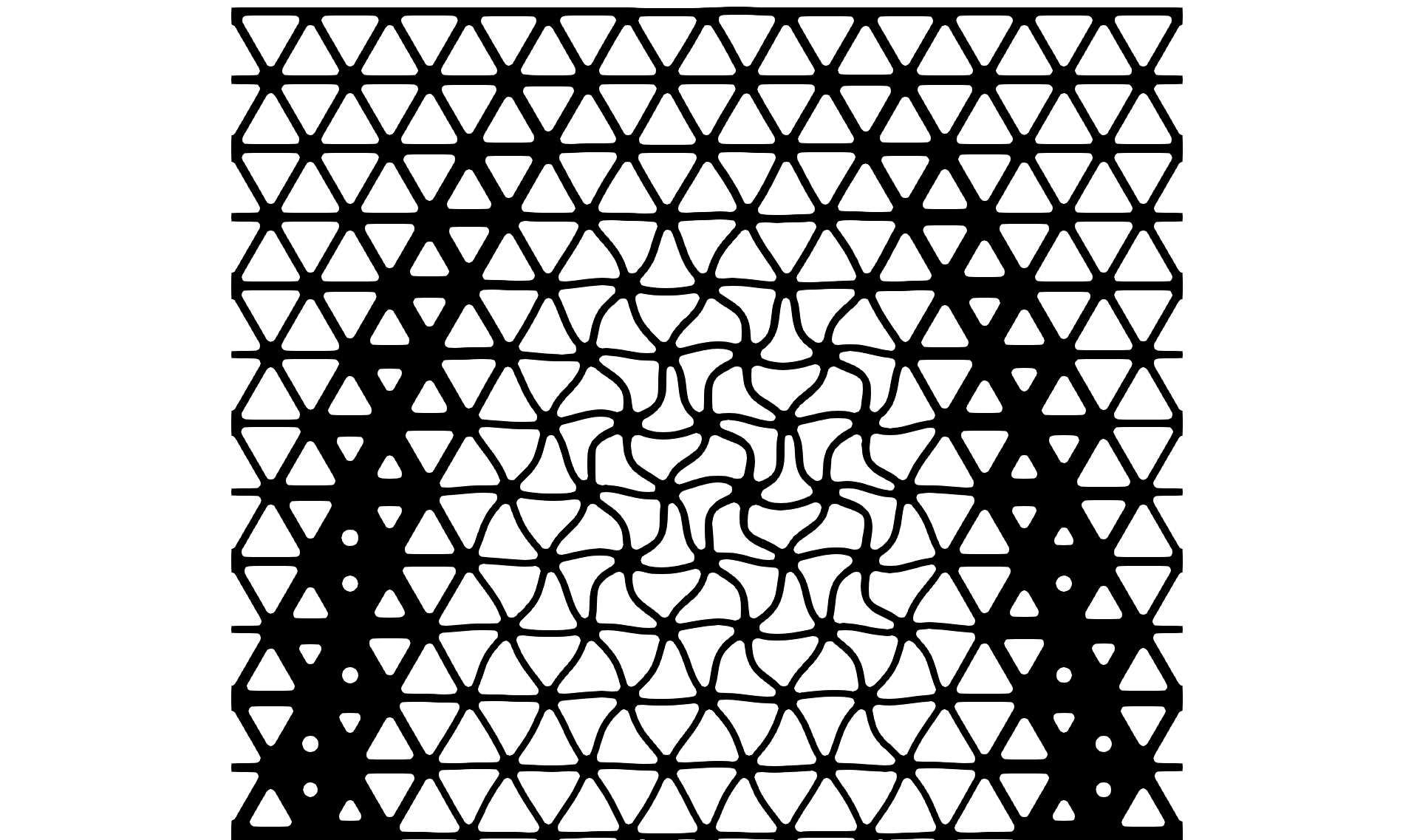}}
  \newsavebox{\valpfemmm}
  \savebox{\valpfemmm}{\includegraphics[trim={300px 0 300px 0},clip,width=.32\columnwidth]{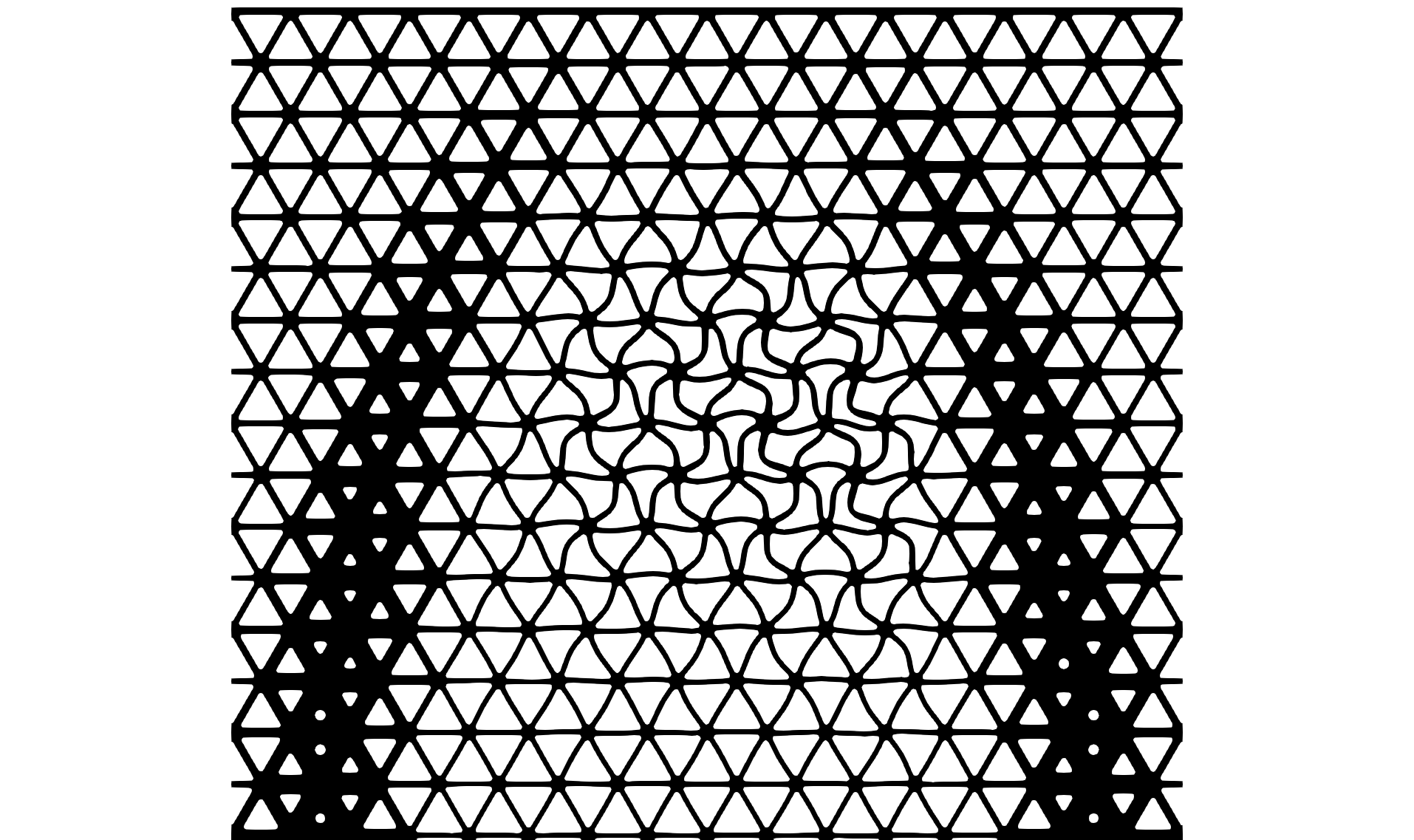}}
  \begin{tikzpicture}
    \node[above right,inner sep=0pt] at (0,0) {\usebox{\valpfem}};
    \fill[white,path fading=north] (0,0) rectangle (\wd\valpfem,.2\ht\valpfem);
  \end{tikzpicture}
  \begin{tikzpicture}
    \node[above right,inner sep=0pt] at (0,0) {\usebox{\valpfemm}};
    \fill[white,path fading=north] (0,0) rectangle (\wd\valpfemm,.2\ht\valpfemm);
  \end{tikzpicture}
  \begin{tikzpicture}
    \node[above right,inner sep=0pt] at (0,0) {\usebox{\valpfemmm}};
    \fill[white,path fading=north] (0,0) rectangle (\wd\valpfemmm,.2\ht\valpfemmm);
  \end{tikzpicture}
  \caption{Zoom of first interior buckling mode of design C$_4$ dehomogenized with $8$ (left), $12$ (middle) and $16$ (right) lattice cells (\cf \cref{fig:cellsize}).}
  \label{fig:dehombucklingzoom}
\end{figure}

\subsubsection*{Comparison between homogenized and dehomogenized designs}
Next, we compare the homogenized design $C_4$ with its dehomogenized counterpart with $20$ cells in a horizontal direction.
For this design, we have a clear separation of length scales and only a small homogenization error.
The macroscopic load factor is determined to be $0.0507$ and differs by less than $1\%$ from the predicted macroscopic load factor of $0.0511$ in the optimized homogenized design.
A gap of $10\%$ exists between the interior microscopic load factor of the dehomogenized design of $0.0563$ and the predicted one of $0.0511$.
This is expected, as the predicted microscopic load factor is based on our worst-case model, which assumes the worst stress type and orientation (see \cref{ssec:worstcasemodel}).
Note that the predicted microscopic buckling load factor is smaller than the microscopic buckling load factor, and hence our worst-case model acts as a safeguard against pure microscopic buckling in this example.
Of course, such an observation is generally only valid up to remaining discretization and interpolation errors as well as the error introduced by homogenization.

\subsection{Impact of worst-case model}\label{ssec:worstcaseimpact}
\begin{figure}[t]
  \centering
  \newsavebox{\lfsim}
  \savebox{\lfsim}{\includegraphics[height=.4\textheight,trim={0 0 106 0},clip]{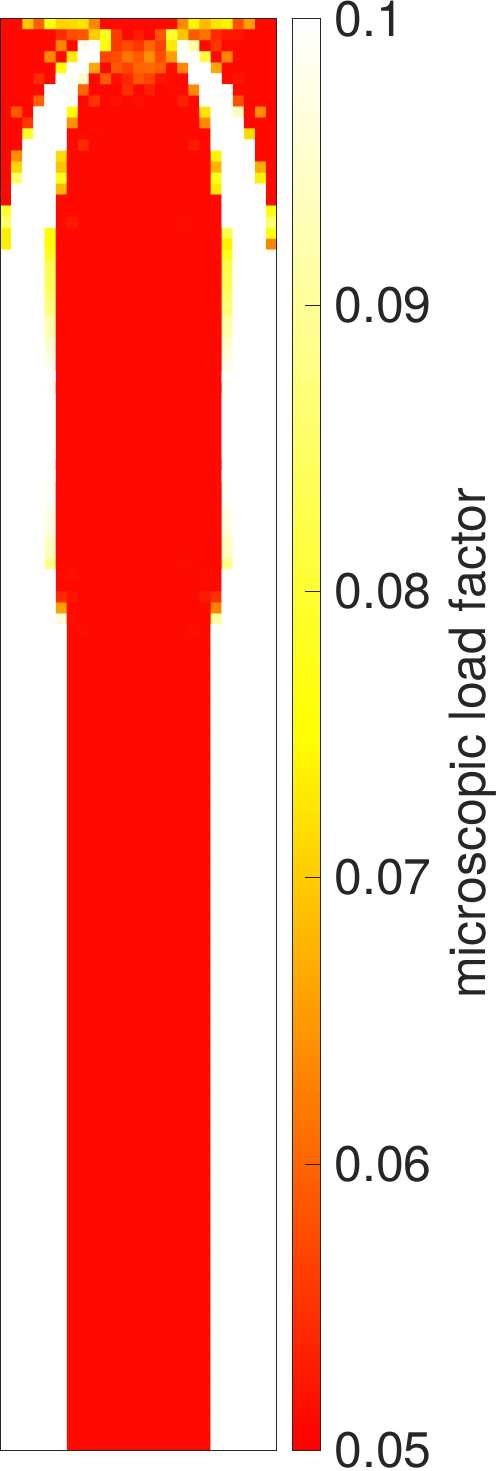}}
  \newsavebox{\lfhom}
  \savebox{\lfhom}{\includegraphics[height=.4\textheight,trim={0 0 101 0},clip]{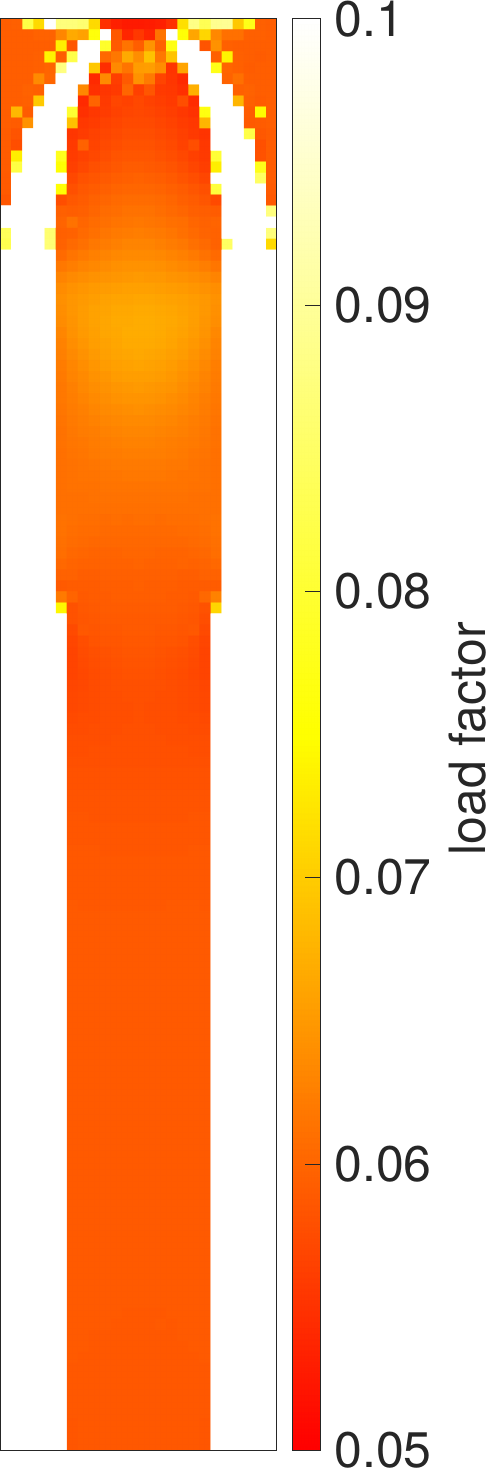}}
  \begin{tikzpicture} [
      square/.style={regular polygon, regular polygon sides=4, inner sep=0pt, thick}
    ]
    \node[above right,inner sep=0pt] at (0,0) {\usebox{\lfsim}};
    \node[square,draw,minimum size=0.08\wd\lfsim] at (.51\wd\lfsim,0.91\ht\lfsim) {};
    \node[above right,inner sep=0pt] at (1.2\wd\lfsim,0) {\usebox{\lfhom}};
    \node[square,draw,minimum size=0.08\wd\lfsim] at (.51\wd\lfhom+1.2\wd\lfsim,0.91\ht\lfhom) {};
  \end{tikzpicture}\qquad
  \includegraphics[height=.4\textheight,trim={140 0 0 0},clip]{loadfactors_hom.png}
  \caption{Left: microscopic buckling load factor predicted by the worst-case model. Right: visualization of a posteriori homogenized microscopic buckling load factor.
  Load factors are thresholded at $0.1$.
  The marked finite element in the upper region corresponds roughly to the center of the mode from \cref{fig:dehombucklingzoom}.
  The a posteriori homogenized buckling load factor of this element is $13\%$ larger than predicted and differs less than $3\%$ from the dehomogenized design.}
  \label{fig:loadfactorshom}
\end{figure}

Next, we investigate the gap between the load factor predicted by the worst-case model and the interior load factor obtained from the dehomogenized design.
For all finite elements in the lattice region of design C$_4$, we perform a posteriori homogenization by evaluation of \cref{eq:fullmodel}.
We extract the local relative volume and create RVEs with $\kappa \in \{4,5,6,7\}$ unit cell repetitions. 
Then, we conduct homogenization (\cref{sec:homogenization}) of the buckling load factor for each $\kappa$ with the "real" macroscopic stress taken from the optimized design using \cref{eq:stressdistribution,eq:macroconsteq} and use the minimal value with respect to $\kappa$.
The resulting microscopic load factors are shown in \cref{fig:loadfactorshom}.
The mean relative difference between predicted and a posteriori load factor for all of the validated finite elements is $18\%$, with a standard deviation of $0.079$.

The marked element in \cref{fig:loadfactorshom} complies with the center of the interior microscopic mode in the dehomogenized design.
For this element, the a posteriori homogenized buckling load factor is $0.0578$, which differs less than $3\%$ from the interior microscopic load factor of the dehomogenized design.
This rather small difference demonstrates that homogenization is a valid tool for predicting microscopic buckling behavior.

The a posteriori homogenized buckling load factor for the mentioned element is $13\%$ larger than predicted by the worst-case model.
This can be explained by \cref{fig:homogenizedlf}.
The macroscopic stress for the chosen finite element, which is extracted from the macroscopic analysis, is uniaxial and acts in vertical direction.
Thus, the a posteriori load factor corresponds to the point on the uniaxial curve at $90^\circ$, which is $15\%$ larger than the biaxial value.

\subsubsection*{Example with tension, compression, and shear stress}
\begin{figure}[t]
  \centering
  \includegraphics[width=.9\columnwidth]{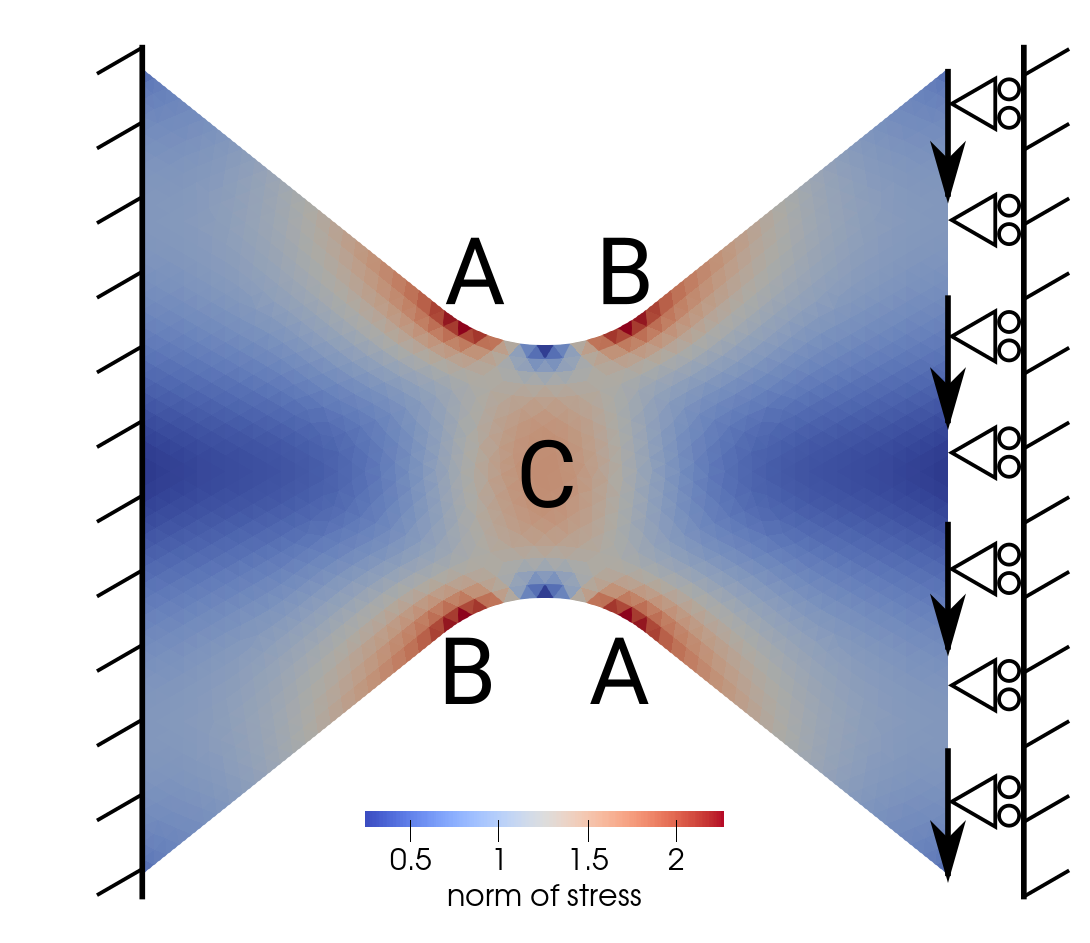}
  \caption{Setting for shear stress example, consisting of homogenized lattice with $50\%$ porosity. The color shows the norm of the resulting mechanical stress. Region C exhibits shear, region A uniaxial tension, and region B uniaxial compression stress.}
  \label{fig:setting_shear}
\end{figure}

In the previously considered example, we mostly observe uniaxial compression stress.
Next, we want to examine an example with compression, tension, and shear stress.
For this, consider the setting shown in \cref{fig:setting_shear}:
A bow-tie shaped domain is fixed at the left edge in both degrees of freedom.
At the right edge, movement in a horizontal direction is prevented and a force is pulling downwards.
We simulate the behavior of a homogenized lattice with a constant relative local volume, i.e., $\rho_e=0.5$ for all $e=1,\ldots,\text{M}$.
This leads to pure shear stress in the center of the domain, marked by C, uniaxial tension stress in the regions marked with A, and uniaxial compression stress in regions marked with B.

\begin{figure}[t]
  \centering
  \newsavebox{\lfshear}
  \savebox{\lfshear}{\includegraphics[width=.9\columnwidth]{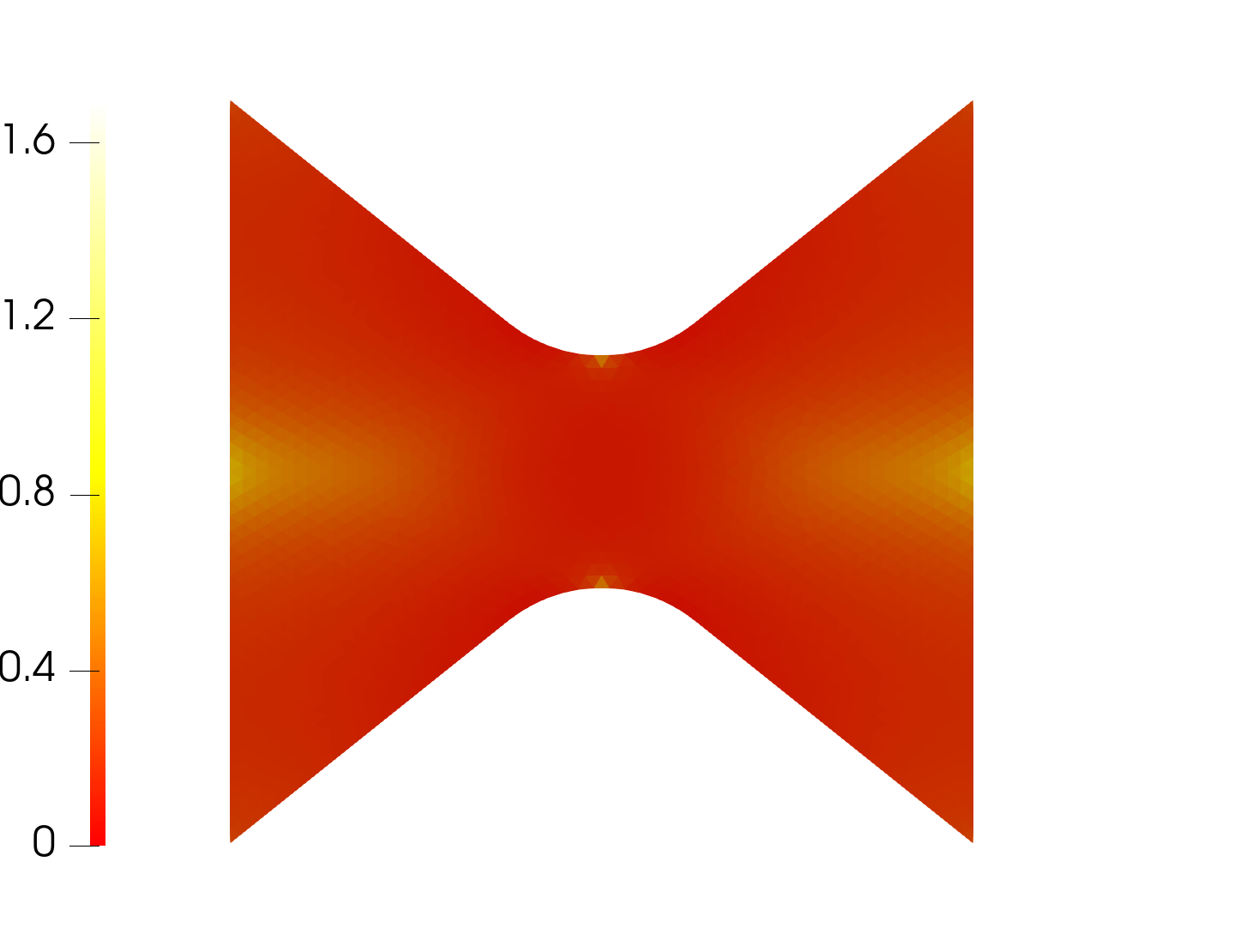}}
  \begin{tikzpicture} [
      dot/.style = {circle, thick, minimum size=#1, inner sep=0pt, outer sep=0pt},
    ]
    \node[above right,inner sep=0pt] at (0,0) {\usebox{\lfshear}};
    \node[dot=0.1\wd\lfshear,draw,anchor=mid] at (.49\wd\lfshear,.5\ht\lfshear) {\small C};
    \node[anchor=mid] at (.42\wd\lfshear,.7\ht\lfshear) {A};
    \node[anchor=mid] at (.56\wd\lfshear,.3\ht\lfshear) {A};
    \node[anchor=mid] at (.42\wd\lfshear,.3\ht\lfshear) {B};
    \node[anchor=mid] at (.56\wd\lfshear,.7\ht\lfshear) {B};
  \end{tikzpicture}
  \caption{Microscopic buckling load factors obtained via our worst-case model.}
  \label{fig:wclf_shear}
\end{figure}
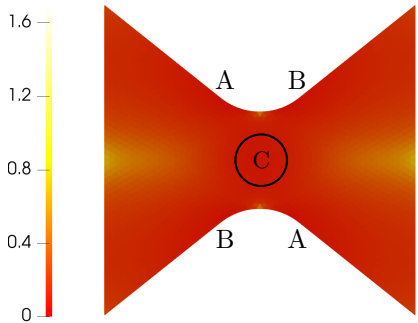

In \cref{fig:wclf_shear} we see, that the microscopic buckling load factor, which is obtained from our worst-case model, has its minimal value in regions A and B where the highest stress occurs.
This highlights a drawback of the worst-case approach:
the model cannot distinguish between tension and compression stress.
This is because it is based on a minimization over all unit stresses and only depends on the local volume fraction and local stress magnitude but not on stress type or direction (\cf \cref{eq:worstcase}).
As for our exemplary lattice, the computed load factors for compression stress are consistently smaller than for tension stress (see also \citep[Fig. 3]{christensen2022multiscale} for a confirmation of this observation), and buckling load factors will always be underestimated by the model in regions of tension stress.
This results in higher lattice volume fraction than actually necessary to prevent local buckling at these locations.
Due to this expendable lattice over-sizing, microscopic buckling is not observed at the corresponding locations for the dehomogenized design, although predicted by the worst-case model.
%does not occur in the dehomogenized design at all locations, which have been predicted to be subject to buckling by the worst-case model.

As in the previous example, we conduct a posteriori homogenization with the "real" macroscopic stress taken from the simulated design in \cref{fig:setting_shear}.
\cref{fig:err_shear} shows that in the region of compression stress (B), the error of the prediction is quite low, but in the region of tension stress (A), the worst-case model clearly underestimates the actual load factor.
For an exemplary finite element in the marked region C, the predicted buckling load factor is $0.100$, while the homogenized value is $0.154$.
Thus, this element actually has $50\%$ higher buckling strength than predicted by the worst-case model.
This can be explained by \cref{fig:homogenizedlf2}.
The load factor for shear stress at $45^\circ$, which is the acting stress in the inspected element, is around $50\%$ larger than the load factor for biaxial stress, on which our worst-case model is based.
This explanation also leads to a simple a priori error estimate for the worst-case model:
if we know the type of stress, the error is bounded by the difference between the maximal load factor on the curve for the specific stress type and the load factor for biaxial stress.

\begin{figure}[t]
  \centering
  \newsavebox{\errshear}
  \savebox{\errshear}{\includegraphics[width=.9\columnwidth]{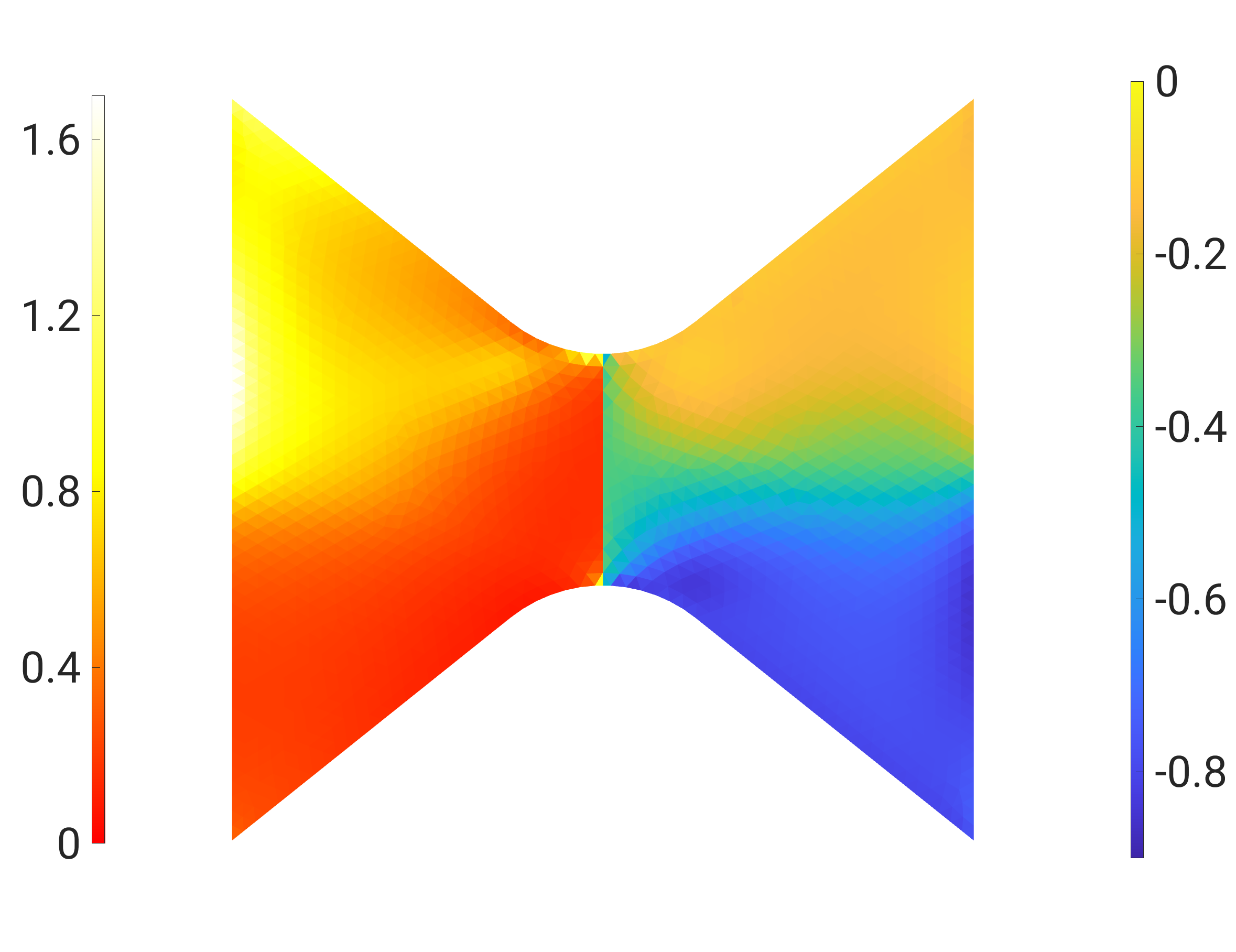}}
  \begin{tikzpicture} [
      dot/.style = {circle, thick, minimum size=#1, inner sep=0pt, outer sep=0pt},
    ]
    \node[above right,inner sep=0pt] at (0,0) {\usebox{\errshear}};
    \node[dot=0.1\wd\errshear,draw,anchor=mid] at (.49\wd\errshear,.5\ht\errshear) {\footnotesize C};
    \node[anchor=mid] at (.42\wd\errshear,.7\ht\errshear) {A};
    \node[anchor=mid] at (.56\wd\errshear,.3\ht\errshear) {A};
    \node[anchor=mid] at (.42\wd\errshear,.3\ht\errshear) {B};
    \node[anchor=mid] at (.56\wd\errshear,.7\ht\errshear) {B};
    \node[anchor=south,inner sep=0pt,rotate=90] at (0,.5\ht\errshear) {a posteriori load factor / -};
    \node[anchor=north,inner sep=0pt,rotate=90] at (\wd\errshear,.5\ht\errshear) {relative error / -};
  \end{tikzpicture}
  \caption{Left: microscopic buckling load factors obtained by a posteriori homogenization with the ``real'' macroscopic stress. Right: relative difference of load factors predicted by the worst-case model compared with a posteriori load factors. The center region C is subject to pure shear stress and has a predicted load factor, which is $50\%$ smaller than the one obtained by a posteriori homogenization.}
  \label{fig:err_shear}
\end{figure}
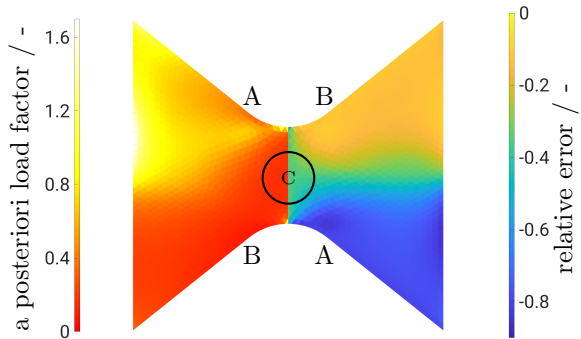

\begin{figure}[t]
  \begin{tikzpicture}
    \pgfplotsset{width=.9\columnwidth}
    \begin{polaraxis}[
      xticklabel={$\pgfmathprintnumber\tick^\circ$},
      xticklabel style={yshift={(\ticknum==1)*.6em}},
      ymin=0,
      ytick distance = 0.05,
      yticklabel style={/pgf/number format/.cd,fixed,fixed zerofill,precision=2},
      extra x ticks={0},
      extra x tick labels={x},
      extra x tick style={tick label style={yshift=-.5em,xshift=.5em}},
      xminorgrids,
      clip=false,
      legend style={inner xsep=1pt, inner ysep=1pt, at={(1.27,1.15)}, anchor=north east}]
      \addplot[black,domain=-180:180,samples=90,line width=1pt] (x,0.0263078985021975);
      \addplot[blue,mark=star,legend image post style={only marks}] table [x expr=\thisrowno{14}*5, y index=0, header=false, only marks] {\uniaxialc};
      \addplot[forget plot,blue,mark=star] table [x expr=\thisrowno{14}*5+ 60, y index=0, header=false, only marks] {\uniaxialc};
      \addplot[forget plot,blue,mark=star] table [x expr=\thisrowno{14}*5+120, y index=0, header=false, only marks] {\uniaxialc};
      \addplot[forget plot,blue,mark=star] table [x expr=\thisrowno{14}*5+180, y index=0, header=false, only marks] {\uniaxialc};
      \addplot[forget plot,blue,mark=star] table [x expr=\thisrowno{14}*5+240, y index=0, header=false, only marks] {\uniaxialc};
      \addplot[forget plot,blue,mark=star] table [x expr=\thisrowno{14}*5+300, y index=0, header=false, only marks] {\uniaxialc};
      \addplot[red,mark=triangle*,legend image post style={only marks}] table [x expr=\thisrowno{14}*5, y index=0, header=false, only marks] {\shear};
      \addplot[forget plot,red,mark=triangle*] table [x expr=\thisrowno{14}*5+ 60, y index=0, header=false, only marks] {\shear};
      \addplot[forget plot,red,mark=triangle*] table [x expr=\thisrowno{14}*5+120, y index=0, header=false, only marks] {\shear};
      \addplot[forget plot,red,mark=triangle*] table [x expr=\thisrowno{14}*5+180, y index=0, header=false, only marks] {\shear};
      \addplot[forget plot,red,mark=triangle*] table [x expr=\thisrowno{14}*5+240, y index=0, header=false, only marks] {\shear};
      \addplot[forget plot,red,mark=triangle*] table [x expr=\thisrowno{14}*5+300, y index=0, header=false, only marks] {\shear};
      \addplot[forget plot,red,mark=triangle*] table [x expr=\thisrowno{14}*5+ 60, y index=0, header=false, only marks] {\shear};
      
      \pgfplotstablegetelem{9}{[index]0}\of{\shear} \pgfmathsetmacro\shearelement{\pgfplotsretval}
      \addplot[forget plot,black,mark=o,mark size=4pt] (45,\shearelement);
      \legend{biaxial compression,uniaxial compression,shear};
      \node (origin) at (axis cs:0,0){};
      \node (xaxis) at (axis cs:0,0.075){};
      \draw[-angle 45] (origin) -- (xaxis);
    \end{polaraxis}
  \end{tikzpicture}
  \hfill
  \caption{Homogenized buckling load factors for an RVE with 3x3 unit cell repetitions and a relative volume of $30\%$ (\cf \cref{fig:homogenizedlf}). In region C of our example (\cref{fig:setting_shear}), we have shear stress acting at an angle of $45^\circ$. The buckling load factor for this stress situation (marked by a circle) is around $50\%$ larger than the one for biaxial compression stress.}
  \label{fig:homogenizedlf2}
\end{figure}
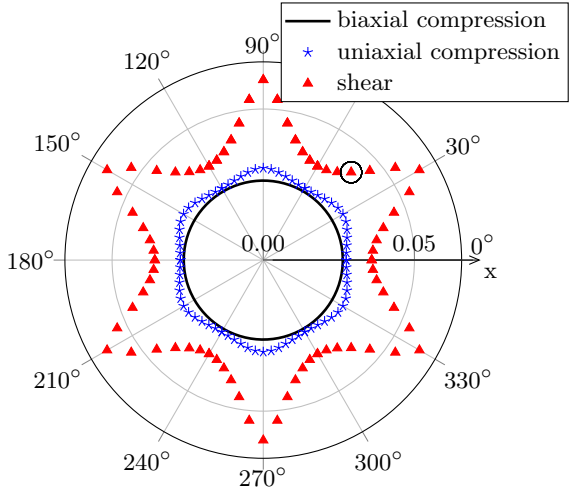

For this last example, we showed a homogeneous design with non-homogeneous microscopic buckling load factor.
In a simultaneous optimization of both macro- and microscopic buckling (\cf problem C in \cref{sec:sizingopt}), only the microscopic constraint is active as the structure exhibits a comparatively high macroscopic buckling load factor.
Optimization with respect to only microscopic buckling (problem B) leads to non-homogeneous lattice with homogeneous microscopic buckling load factor.
For this optimized design, similar deviations in the predicted worst case versus the a posteriori buckling load factor have been found in regions that exhibit shear stress.

We recall that our worst-case model acts as a safeguard against pure microscopic buckling.
The possibly excessive underestimation of the microscopic buckling load factor can lead to oversized lattice struts.
This can be overcome by replacing the worst-case approach with other interpolation models, e.g., a $C^1$-Interpolation in the three-dimensional parameter space $(\rho,\vec{\bar\sigma})|_{\vec{\bar\sigma}\in S^2}$.
However, this comes with additional computational effort for the worst-case model and, as mentioned in \cref{rem:wcfree} in \cref{ssec:worstcasemodel}, requires sophisticated interpolation schemes if the proposed method is applied for three-dimensional structures, as the parameter space will become higher dimensional.

\section{Conclusion}\label{sec:conclusion}%
We presented a method to perform two-scale optimization of lattice structures while accounting for buckling on both scales using asymptotic homogenization.
Based on a parameterization of our chosen lattice structure, we obtained homogenized elastic and buckling properties.
We constructed a worst-case model for the homogenized buckling load factor (\cref{ssec:worstcasemodel}).
Both elastic and buckling characteristics were upscaled by individual, continuously differentiable interpolation models.
We provided numerical examples for optimization of only macroscopic or microscopic buckling and simultaneous optimization of both.

We demonstrated that our method to obtain homogenized properties is equivalent to a special discretization of the Brillouin zone in Floquet-Bloch theory.
We showed that rounding corners can have a considerable influence on the buckling strength of lattices, as it prevents stress concentrations.

We compared the performance of optimized designs predicted by the worst-case model with their dehomogenized counterparts.
Compliance and macroscopic buckling were predicted very well; the predicted microscopic buckling load factor was conservative.
We explained the reason for a possible underestimation and how this can be avoided by replacing the worst-case model by interpolation in the whole parameter space.
A posteriori homogenization matched dehomogenized results quite well.
This shows that homogenization is a viable tool to upscale the buckling behavior of lattice structures.

Although we limit ourselves to a two-dimensional setting in this article, an extension to three dimensions is straight forward.
Further research could include the combination of lattice, solid, and void design in optimization problems while accounting for manufacturing constraints.

\backmatter

\bmhead{Acknowledgments}

Funded by the Deutsche Forschungsgemeinschaft (DFG, German Research Foundation) – Project-ID 61375930 – SFB 814 - "Additive Manufacturing" TP C02.
\section*{Declarations}%
\begin{description}
\item [Competing interests]
On behalf of all authors, the corresponding author states that there is no conflict of interest.
\item [Replication of results]
The numerical implementation of the presented approach is based on and extends the open-source finite element package \emph{openCFS} (\citeauthor{openCFS}).
The presented numerical examples can be reproduced following the general \emph{openCFS} instructions.
Examples on how to conduct homogenization and buckling analysis are provided in its \emph{Testsuite}.
Finite element meshes have been generated by \emph{Coreform Cubit 2022.4} (\citeauthor{cubit}).
Additional scripts for post-processing of the homogenized data are readily provided by the authors on request.
\end{description}


%\bibliographystyle{sn-vancouver}
\bibliography{literature}


\end{document}